\documentclass[12pt,twoside,english]{article}
\usepackage{bookman}
\usepackage[T1]{fontenc}
\usepackage[latin1]{inputenc}
\usepackage{amsmath}
\usepackage{graphicx}
\usepackage{amssymb}

\makeatletter

\providecommand{\LyX}{L\kern-.1667em\lower.25em\hbox{Y}\kern-.125emX\@}

 \usepackage{verbatim}

\usepackage{bookman}
\usepackage{a4}
\usepackage{cite}
\input{epsfig.sty}
\def\fnum@table{\tablename~{\bf\thetable}}
\def\fnum@figure{\figurename~{\bf\thefigure}}
\def\tablename{\footnotesize{\bf Table}}
\def\figurename{\footnotesize{\bf Figure}}

\AtBeginDocument{

}

\usepackage{babel}
\makeatother
\begin{document}
\begin{center}\textbf{\huge Parton Ladder Splitting }\end{center}{\huge \par}

\begin{center}\textbf{\huge and the Rapidity Dependence }\end{center}{\huge \par}

\begin{center}\textbf{\huge of Transverse Momentum Spectra }\end{center}{\huge \par}

\begin{center}\textbf{\huge in Deuteron-Gold Collisions at RHIC}\end{center}{\huge \par}

\vspace{0.6cm}
\begin{center}\textbf{\large Klaus WERNER} {\large $^{1}$}\textbf{\large ,
Fu-Ming LIU $^{2}$, Tanguy PIEROG} \textbf{\Large $^{3}$}\end{center}{\Large \par}
\vspace{0.6cm}

\noindent \begin{center}\textit{$^{1}$ SUBATECH, Université de Nantes
-- IN2P3/CNRS-- EMN,  Nantes, France}\end{center}

\noindent \begin{center}\textit{\emph{$^{2}$}} \emph{Central China
Normal University, Institute of Particle Physics, Wuhan, China}\end{center}

\noindent \begin{center}\emph{$^{3}$ Forschungszentrum Karlsruhe,
Institut f. Kernphysik, Karlsruhe, Germany}\end{center}
\vspace{0.5cm}

\noindent \textbf{Abstract}: We present a phenomenological approach
(EPOS), based on the parton model, but going much beyond, and try
to understand proton-proton and deuteron-gold collisions, in particular
the transverse momentum results from all the four RHIC experiments.
It turns out that elastic and inelastic parton ladder splitting is
the key issue. Elastic splitting is in fact related to screening and
saturation, but much more important is the inelastic contribution,
being crucial to understand the data. We investigate in detail the
rapidity dependence of nuclear effects, which is actually relatively
weak in the model, in perfect agreement with the data, if the latter
ones are interpreted correctly.

\section{Introduction}

Interesting new results have been observed in heavy ion collisions
at RHIC, a large fraction of which being based on transverse momentum
spectra. High transverse momenta seem to be suppressed \cite{star-aa,phenix-aa},
systematically different for different hadron species \cite{phenix-aa-2}. 

But any quantitative expression of a suppression or an enhancement
needs a reference, and here one usually refers to proton-proton scattering.
Rather than working with the spectrum, one investigates the ratio
$R_{AA}$ of the nucleus-nucleus ($AA$) spectrum to the proton-proton
($pp$) result, the so-called nuclear modification factor,\begin{equation}
R_{AA}=\frac{1}{N_{\mathrm{coll}}}\, \frac{dn^{AA}}{d^{2}p_{t}\, dy}\, /\, \frac{dn^{pp}}{d^{2}p_{t}\, dy}\, .\end{equation}
The normalization factor $1/N_{\mathrm{coll}}$ has been chosen such
that at large transverse momenta one expects $R_{AA}$ to become unity.
Here, $dn$ refers to the number of produced particles per inelastic
interaction. 

The first problem is therefore to understand sufficiently well proton-proton
scattering. This is far from trivial. Experimentally, it is difficult
to really access the full inelastic cross section, the interaction
triggers tend to miss a more or less large fraction of the events.
Theoretically, proton-proton is far from being fully understood, apart
from perturbative calculations concerning very large transverse momenta.

A second problem arises due to the fact that even being sure about
the observation of a non-trivial behavior of $R_{AA}$, we want to
know whether this effect is really a collective one, providing evidence
of the formation of a quark gluon plasma, and not something we observe
already in proton-nucleus. This was the main purpose to study, in
addition to gold-gold ($AuAu$), as well deuteron-gold ($dAu$) collisions
at RHIC \cite{star-a,star-b,phenix-a,phobos-b,brahms-b}, with quite
interesting results: the strong high $p_{t}$ suppression in $AuAu$
seems to be absent in $dAu$, so we have clearly a final state effect. 

Many features of $dAu$ seem to be qualitatively understood, employing
the saturation model \cite{saturation1,saturation2,saturation3,saturation4},
a recombination model \cite{recombination}, an improved parton model
\cite{partonmodel1,partonmodel2,partonmodel3}, the AMPT model\cite{AMPT}.
But, what is really missing is a global and quantitative investigation:
can we understand ALL the data presented so far by ALL the experiments,
for $pp$ and $dAu$, in a single approach. This gives also the opportunity
to cross-check the different experiments, which is not so obvious
to do directly.

The purpose of this paper is to present a phenomenological approach
(EPOS), based on the parton model, but going much beyond, and try
to understand $pp$ and $dAu$, as far as the transverse momentum
results from all the four RHIC experiments are concerned. It turns
out that parton splitting (or better parton ladder splitting) is the
key issue, which is related to screening and saturation, but there
are other important consequences, which are crucial to understand
the data.

\section{Improved Parton Model with Remnants}

The new approach we are going to present is called EPOS, which stands
for

\begin{itemize}
\item {\Large E}nergy conserving quantum mechanical multiple scattering
approach, \\
$\, $\hspace*{1cm}based on 
\item {\Large P}artons (parton ladders)
\item {\Large O}ff-shell remnants
\item {\Large S}plitting of parton ladders
\end{itemize}
We are going to explain the different items in the following (the
parton splitting will be discussed in a later section).

\begin{figure}
\begin{center}\includegraphics[  scale=0.8]{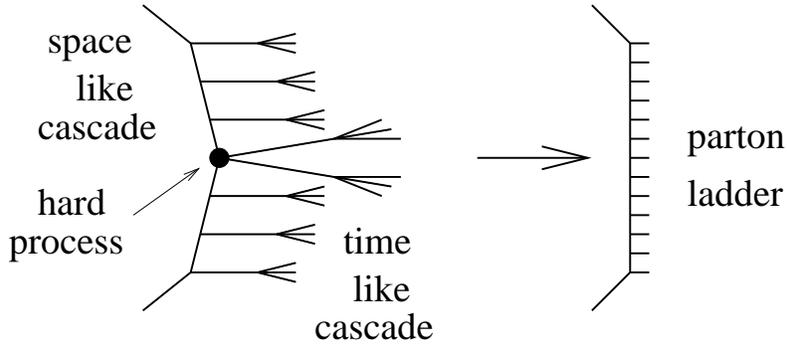}\end{center}

\caption{Elementary parton-parton scattering: the hard scattering in the middle
is preceded by parton emissions (initial state space-like cascade);
these partons being usually off-shell, they emit further partons (final
state time-like cascade). For all this we use a symbolic parton ladder.
\label{ladder} }
\end{figure}

One may consider the simple parton model to be the basis of hadron-hadron
interaction models at high energies. It is well known that the inclusive
cross section is given as a convolution of two parton distribution
functions with an elementary parton-parton interaction cross section.
The latter one is obtained from perturbative QCD, the parton distributions
are deduced from deep inelastic scattering. Although these distributions
are taken as black boxes, one should not forget that they represent
a dynamical process, namely the successive emission of partons (initial
state space-like cascade), which have to be considered in a complete
picture. In addition, the produced partons are generally off-shell,
giving rise again to parton emissions (final state time-like cascade).
All this is sketched in fig. \ref{ladder}, where we also indicate
that we refer to this whole structure as {}``parton ladder'', with
a corresponding simple symbol, to simplify further discussion.

For practical calculations, each parton ladder is finally translated
into two color strings, which fragment into hadrons. This is a purely
phenomenological procedure for the non-perturbative hadronization
process.

Actually our {}``parton ladder'' is meant to contain two parts:
the hard one, as discussed above, and a soft one, which is a purely
phenomenological object, parametrized in Regge pole fashion, for details
see Appendix \ref{Appendix-soft-hard}.

Still the picture is not complete, since so far we just considered
two interacting partons, one from the projectile and one from the
target. These partons leave behind a projectile and target remnant,
colored, so it is more complicated than simply projectile/target deceleration.
One may simply consider the remnants to be diquarks, providing a string
end, but this simple picture seems to be excluded from strange antibaryon
results at the SPS \cite{sbaryons}. 

\begin{figure}
\begin{center}\includegraphics[  scale=0.8]{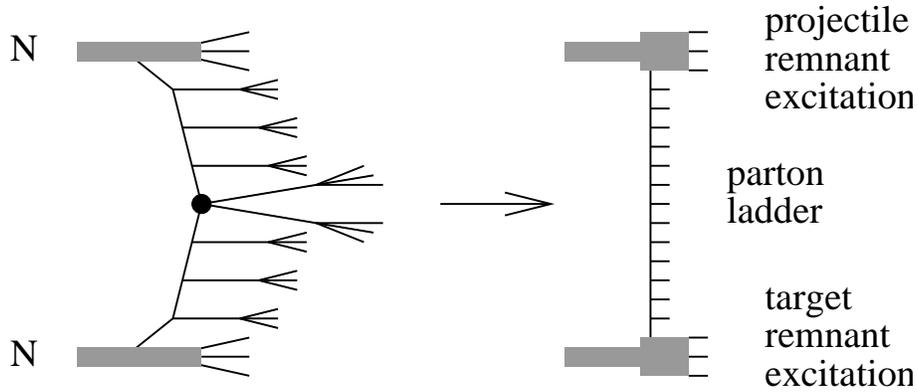}\end{center}

\caption{The complete picture, including remnants. The remnants are an important
source of particle production at RHIC energies.\label{complete}}
\end{figure}

\begin{figure}
\begin{center}\includegraphics[  scale=0.8]{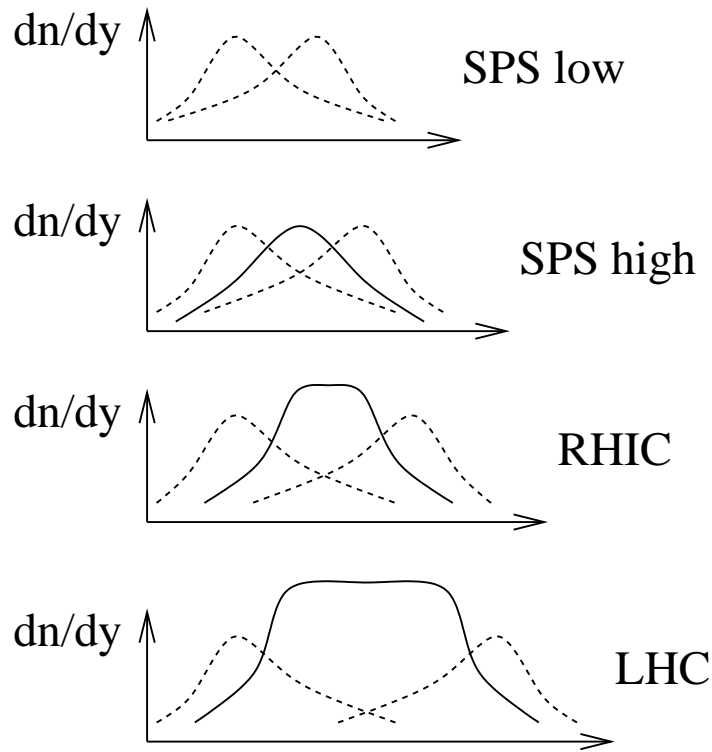}\end{center}

\caption{Inner contributions, from the parton ladder (full lines), and {}``outer''
contributions, from the remnants (dashed lines), to the rapidity distribution
of hadrons. (Artists view)\label{inout} }
\end{figure}

We therefore adopt the following picture, as indicated in fig. \ref{complete}:
not only a quark, but a two-fold object takes directly part in the
interaction, being a quark-antiquark, or a quark-diquark, leaving
behind a colorless remnant, which is, however, in general excited
(off-shell). So we have finally three {}``objects'', all being white:
the two off-shell remnants, and the parton ladders between the two
active {}``partons'' on either side (by {}``parton'' we mean quark,
antiquark, diquark, or antidiquark). We also refer to {}``inner contributions''
(from parton ladders) and {}``outer contributions'' (from remnants),
reflecting the fact that the remnants produce particles mainly at
large rapidities and the parton ladders at central rapidities, see
fig. \ref{inout}.%
Whereas the outer contributions are essentially energy independent,
apart of a shift in rapidity, the inner contributions grows with energy,
to eventually dominate completely central rapidities. But at RHIC,
there is still a substantial remnant contribution at mid-rapidity.

Even inclusive measurements require often more information than just
inclusive cross sections, for example via trigger conditions. Anyhow,
for detailed comparisons we need an event generator, which obviously
requires information about exclusive cross sections (the widely used
pQCD generators are not event generators in this sense, they are generators
of inclusive spectra, and a Monte Carlo event is not a physical event).
This problem is known since many years, the solution is Gribov's multiple
scattering theory, employed since by many authors. This formulation
is equivalent to using the eikonal formula to obtain exclusive cross
sections from the knowledge of the inclusive one. 

We indicated recently inconsistencies in this approach, proposing
an {}``energy conserving multiple scattering treatment''. The main
idea is simple: in case of multiple scattering, when it comes to calculating
partial cross sections for double, triple ... scattering, one has
to explicitly care about the fact that the total energy has to be
shared among the individual elementary interactions.

\begin{figure}
\begin{center}\includegraphics[  scale=0.8]{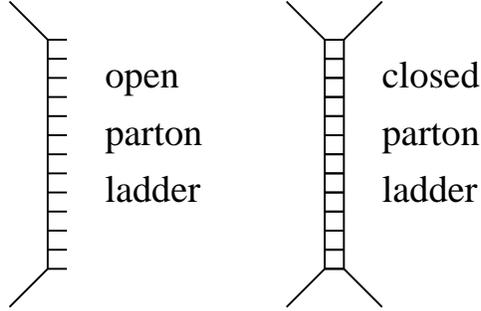}\end{center}

\caption{The two elements of the multiple scattering theory: open ladders,
representing inelastic interactions, and closed ladders, representing
elastic interactions.\label{openclosedladder}}
\end{figure}

A consistent quantum mechanical formulation requires not only the
consideration of the (open) parton ladders, discussed so far, but
also of closed ladders, representing elastic scattering, see fig.
\ref{openclosedladder}. %
The closed ladders do not contribute to particle production, but they
are crucial since they affect substantially the calculations of partial
cross sections. Actually, the closed ladders simply lead to large
numbers of interfering contributions for the same final state, all
of which have to be summed up to obtain the corresponding partial
cross sections. For details see appendix \ref{Appendix-Multiple}.

We can do the complicated calculations, since we fit for example the
result of a numerical calculation of a squared amplitude corresponding
to a (open) parton ladder of energy $\sqrt{s}$, using a simple form
$\alpha s^{\beta }$, which allows then to perform analytical calculations.
Furthermore, we employ very sophisticated Markov chain techniques
to generate configurations according to multidimensional probability
distributions. For details see appendix \ref{Appendix-Parameterizations}
and appendix \ref{Appendix-Markov}.

Important concerning numerical results: There are a couple of parameters
which determine the parameterization of the soft elementary interaction
(soft Pomeron), which are essentially fixed to get the $pp$ cross
sections right. The pQCD parameters (soft virtuality cutoff, K-factor,
parton emission cutoff, parton-hadron coupling) are fixed to provide
a reasonable parton distribution function (which we calculate, it
is not input!).

We assume the remnants to be off-shell with probability $p_{O}$,
a mass distribution given as \begin{equation}
\mathrm{prob}\propto M^{-2\alpha _{O}},\end{equation}
with parameter values which are not necessarily the same for diffractive
and nondiffractive interactions (the latter ones being defined to
be those without parton ladders). We use currently for $p_{O}$ 0.75
(dif) and 0.95 (nondif), and for $\alpha _{O}$ 0.75 (dif) and 1.1
(nondif). Those excitation exponents may give rise to quite high mass
remnants, RHIC and also SPS data seem to support this. High mass remnants
will be treated as strings

There are four important fragmentation parameters: the break probability
(per unit space-time area) $p_{B}$, which determines whether a string
breaks earlier or later, the diquark break probability $p_{D}$, the
strange break probability $p_{S}$, and the mean transverse momentum
$\bar{p}_{t}$ of a break, with obvious consequences for baryon and
strangeness production, and the $p_{t}$ of the produced hadrons.
We use three sets of these parameters, for the three types of strings:
soft-, kinky(hard)-, remnant-strings. We do not really use the full
freedom of these parameters, but one single set would not work --
if we are interested in high precision. Somewhat surprising: $p_{S}$
is 0.14 for soft and 0.06 for kinky strings. Maybe this reflects the
fact that soft strings may have low masses, where strangeness is suppressed,
and which needs some compensation. The parameter $p_{D}$ is as well
bigger for soft compared to kinky strings.

\section{Splitting of Parton Ladders}

\begin{figure}
\begin{center}\includegraphics[  scale=0.7]{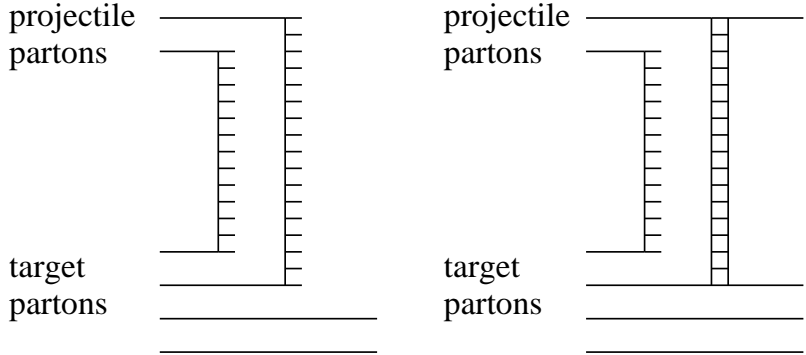}\end{center}

\caption{Basic parton-parton interaction in nucleus-nucleus collisions: a
projectile parton always interacts with exactly one parton on the
other side, elastically (closed parton ladder) or inelastically (open
parton ladder). \label{basic}}
\end{figure}

Let us first consider very asymmetric nucleus-nucleus collisions,
like proton-nucleus or deuteron-nucleus. The formalism developed earlier
for $pp$ can be generalized to these nuclear collisions, as long
as one assumes that a projectile parton always interacts with exactly
one parton on the other side, elastically or inelastically (realized
via closed or open parton ladders), see fig. \ref{basic}.

We employ the same techniques as already developed in the previous
section. The calculations are complicated and require sophisticated
numerical techniques, but they can be done. The corresponding results
for $dAu$ will be discussed later.

\begin{figure}
\begin{center}\includegraphics[  scale=0.7]{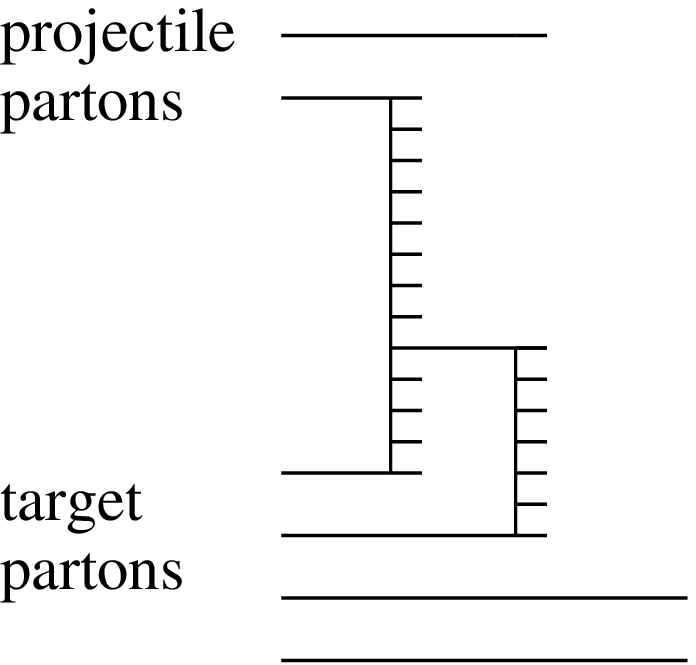}$\qquad $\includegraphics[  scale=0.7]{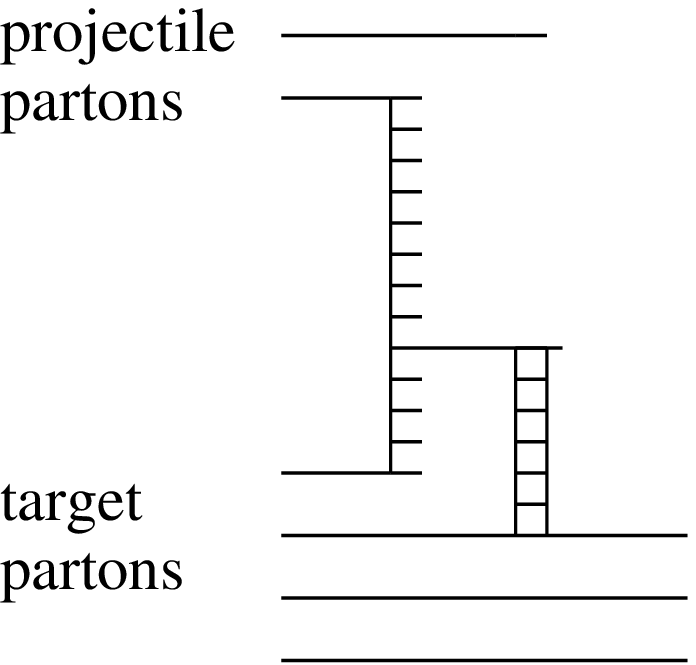}\end{center}

\caption{Inelastic and elastic {}``rescattering'' of a parton from the parton
ladder with a second target parton. We talk about (inelastic and elastic)
splitting of a parton ladder. \label{split}}
\end{figure}

In case of protons (or deuterons) colliding with heavy nuclei (like
gold), there is a complication, which has to be taken into account:
suppose an inelastic interaction involving an open parton ladder,
between a projectile and some target parton. The fact that these two
partons interact implies that they are close in impact parameter (transverse
coordinate). Since we have a heavy target, there are many target partons
available, and among those there is a big chance to find one which
is as well close in impact parameter to the two interacting partons.
In this case it may be quite probable that a parton from the ladder
interacts with this second target parton, inelastically or elastically,
as shown in fig. \ref{split}.

Let us first discuss the effects of elastic splitting. The squared
amplitude for an elementary inelastic interaction involving two partons
with light cone momentum shares $x^{+}=2p^{+}/\sqrt{s}$ and $x^{-}=2p^{-}/\sqrt{s}$
can be parametrized quite accurately as \begin{equation}
\alpha \, (x^{+})^{\beta }(x^{-})^{\beta },\end{equation}
with two parameters $\alpha $ and $\beta $ depending on the squared
energy $s$ and the impact parameter $b$ ($\sqrt{s}$ is the proton-proton
cms energy). Any addition of an elastic contribution (closed ladder),
be it in parallel or via splitting, provides an interference term,
contributing negatively to (partial) cross sections. So an additional
elastic leg, even though it does not affect particle production, it
provides screening. Model calculations show that adding elastic splittings
to the basic diagrams, modifies the corresponding squared amplitude
as\[
\alpha \, (x^{+})^{\beta }(x^{-})^{\beta +\varepsilon },\]
and therefore the whole effect can be summarized by a simple positive
exponent $\varepsilon $, which suppresses small light cone momenta.
So the existence of many target partons effectively screens small
$x$ contributions, which agrees qualitatively with the concept of
saturation. But this is only a part of the whole story, several other
aspects have to be considered.

\begin{figure}
\begin{center}\includegraphics[  scale=0.7]{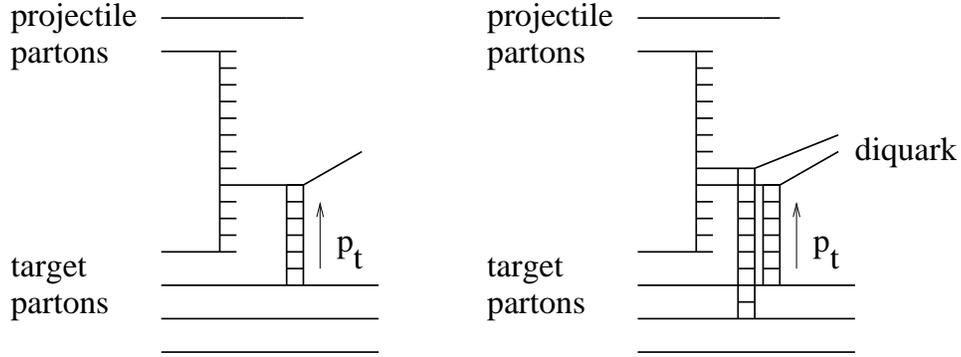}\end{center}

\caption{Transport of transverse momentum via an attached closed ladder, which
may be even enhanced in case of diquarks. \label{trans}}
\end{figure}

One effect is the transport of transverse momentum via an attached
closed ladder, as shown in fig. \ref{trans}.%
Such a transport we use already in the basic parton model, when it
comes to diffractive scattering, realized via a closed ladder. Here,
some transverse momentum transfer is needed to explain the transverse
momentum spectra of protons at large $x$ (in the diffractive region).
In case of diffractive target excitation, the projectile gets simply
a $p_{t}$ kick. We should have the same phenomenon in case of elastic
splitting: the ladder parton involved in the interaction should get
a $p_{t}$ kick in the same way as the proton in diffractive scattering.
This could be even more effective for diquarks (two attached ladders),
leading finally to baryon production.

\begin{figure}
\begin{center}\includegraphics[  scale=0.7]{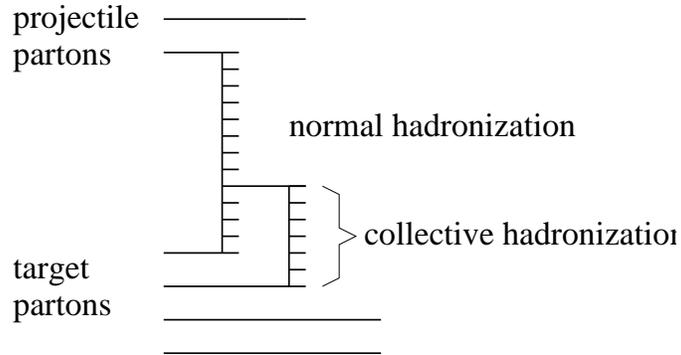}\end{center}

\caption{Hadron production in case of inelastic ladder splitting. \label{splitinel}}
\end{figure}

Let us turn to inelastic splitting, fig. \ref{splitinel} .%
Consider the example shown in the figure. In the upper part, there
is only an ordinary parton ladder, so we expect {}``normal'' hadronization.
In the lower part, we have two ladders in parallel, which are in addition
close in space, since they have a common upper end, and the lower
ends are partons close in impact parameter, so the hadronization of
the two ladders is certainly not independent, we expect some kind
of a {}``collective'' hadronization of two interacting ladders.
Here, we only considered the most simple situation, one may also imagine
three or more close ladders, hadronizing collectively.

So far we discussed in a qualitative fashion the consequences of elastic
and inelastic parton ladder splitting. The strength of the effects
will certainly depend on the target mass, via the number $Z$ of partons
available for additional legs. The number $Z$ of available partons
will also increase with energy, so at high enough energy the above-mentioned
effects can already happen in $pp$ collisions.

\section{Realization of Ladder Splitting Effects\label{sec:Realization-of-Ladder}}

The basic quantity for a numerical treatment of the ladder splitting
effects is the number $Z$ of partons available for additional legs,
more precisely we have a $Z_{T}$ for counting legs on the target
side, and $Z_{P}$ for counting legs on the projectile side. Let us
treat $Z_{T}$ (corresponding discussion for $Z_{P}$). Consider a
parton in projectile nucleon $i$ which interacts with a parton in
target nucleon $j$. The number $Z_{T}(i,j)$ of addition legs has
two contributions, one counting the legs attached to the same nucleon
$j,$ and one counting the legs attached to the other nucleons $j'\neq j$.
We assume the following form:\begin{eqnarray*}
Z_{T}(i,j) & = & z_{0}\, \exp (-b_{ij}^{2}/2b_{0}\! ^{2})+\\
 &  & +\sum _{\mathrm{target}\, \mathrm{nucleons}\, j'\neq j}z'_{0}\, \exp (-b_{ij'}^{2}/2b_{0}\! ^{2}),
\end{eqnarray*}
where $b_{ij}$ is the distance in impact parameter between $i$ and
$j$. The coefficients $z_{0}$ and $z'_{0}$ depend logarithmically
on the energy, as\begin{eqnarray*}
z_{0} & = & w_{Z}\, \ln \, s/s_{M}\, ,\\
z'_{0} & = & w_{Z}\, \sqrt{(\ln \, s/s_{M})^{2}+w_{M}\! ^{2}}\, ,
\end{eqnarray*}
and the impact parameter width is $b_{0}=w_{B}\, \sqrt{\sigma _{\mathrm{inel}\, pp}/\pi }$,
with parameters $w_{B}$, $w_{Z},$ $w_{M}$, and $s_{M}$. We then
define \[
Z_{T}(j)=\sum _{i}Z_{T}(i,j).\]

We suppose that all the effects of the parton ladder splitting can
be treated effectively, meaning that the correct explicit treatment
of splittings is equivalent to the simplified treatment without splittings,
but with certain parameters modified, expressed in terms of $Z$.

This is not only to simplify our life. Even an explicit dynamical
treatment will stay a phenomenological approach, with many uncertainties,
for example about the splitting vertices, and much more. So we prefer
to have simple parameterizations rather than a very complicated but
uncertain dynamical treatment.

So which quantities depend on $Z$, and how? In the following the
symbols $a_{i}$ are constants, used as fit parameters. The elastic
splitting leads to screening, which is expressed by the screening
exponents $\varepsilon =\varepsilon _{S}$ (for soft ladders) and
$\varepsilon =\varepsilon _{H}$ (for hard ladders), and here we assume\begin{equation}
\varepsilon _{S}=a_{S}\, \beta _{S}\, Z\, ,\end{equation}
\begin{equation}
\varepsilon _{H}=a_{H}\, \beta _{H}\, Z\, ,\end{equation}
where $\beta _{S}$ and $\beta _{H}$ are the usual exponents describing
soft and hard amplitudes (see appendix \ref{Appendix-Parameterizations}).

A second effect is transport of transverse momentum, here we suppose\begin{equation}
\Delta p_{t}=a_{T}\, p_{0}\, n_{q}\, Z,\end{equation}
where $n_{q}$ is the number of quarks of the objects in the hadronization
process (1 for quarks, 2 for diquarks), and $p_{0}=0.5$ GeV is just
used to define a scale. 

Let us come to the collective hadronization. We will actually {}``absorb''
the multiple ladders into the remnants, which are usually treated
as strings. Now we treat them as strings with modified string break
parameters, to account for the collective hadronization. We modify
the break probability (per unit space-time area) $p_{B}$, which determines
whether a string breaks earlier or later, the diquark break probability
$p_{D}$, the strange break probability $p_{S}$, and the mean transverse
momentum $\bar{p}_{t}$ of a break, as\begin{eqnarray}
p_{B} & \to  & p_{B}-a_{B}\, Z\, ,\\
p_{D} & \to  & p_{D}\, (1+a_{D}\, Z)\, ,\\
p_{S} & \to  & p_{S}\, (1+a_{S}\, Z)\, ,\\
\bar{p}_{t} & \to  & \bar{p}_{t}\, (1+a_{P}\, Z)\, ,
\end{eqnarray}
with positive parameters $a_{i}$. So with increasing $Z$, a reduced
$p_{B}$ will lead to more particle production, an increased $p_{D}$,
$p_{S}$, $\bar{p}_{t}$, will lead to more baryon-antibaryon production,
more strangeness production, and an increased $p_{t}$ for each string
break. 

The parameters $s_{M}$, $w_{i}$, and $a_{i}$ are chosen to reproduce
mainly RHIC $pp$ and $dAu$ data, but also the energy dependence
of cross sections and multiplicities from SPS to Tevatron. The best
fit parameters are shown in table \ref{param}.%
\begin{table}
\begin{center}\begin{tabular}{|c|l|c|}
\hline 
coefficient&
corresponding variable&
value\\
\hline
\hline 
$s_{M}$&
Minimum squared screening energy&
(25 GeV)$^{2}$\\
\hline 
$w_{M}$&
Defines minimum for $z'_{0}$ &
6.000\\
\hline 
$w_{Z}$&
Global $Z$ coefficient &
0.080\\
\hline 
$w_{B}$&
Impact parameter width coefficient&
1.160\\
\hline 
$a_{S}$&
Soft screening exponent &
2.000\\
\hline 
$a_{H}$&
Hard screening exponent&
1.000\\
\hline 
$a_{T}$&
Transverse momentum transport&
0.025\\
\hline 
$a_{B}$&
Break parameter&
0.070\\
\hline 
$a_{D}$&
Diquark break probability&
0. 110\\
\hline 
$a_{S}$&
Strange break probability&
0.140\\
\hline 
$a_{P}$&
Average break transverse momentum&
0.150\\
\hline
\end{tabular}\end{center}

\caption{Best fit values for splitting parameters. We included in the fit
as well data not shown in this paper. \label{param}}
\end{table}

\section{Results for Proton-Proton}

Ladder splitting is quite important for $pp$ at very high energies,
where cross sections and multiplicities are considerably suppressed,
due to screening. At RHIC energies, however, the effects are small:
the total cross section is reduced by 5\%, the multiplicity by 10\%.
Concerning the transverse momentum spectra to be discussed in detail
in the following, the effect is hardly visible.

In order to compare to the charged particle $p_{t}$ spectra in $pp$
from the different experiments (STAR, PHENIX, BRAHMS), one has first
to understand what has been measured. One wants to measure the inelastic
differential yield,\begin{equation}
\frac{d^{3}n^{\mathrm{inel}}}{dyd^{2}p_{t}}=\frac{1}{\sigma _{\mathrm{inel}}}\, \frac{d^{3}\sigma ^{\mathrm{inel}}}{dyd^{2}p_{t}},\end{equation}
where $\sigma \approx 42$ mb is the inelastic $pp$ cross section,
and $d^{3}\sigma ^{\mathrm{inel}}/dyd^{2}p_{t}$ represents the inclusive
differential cross section for inelastic events. In practice there
is an event trigger like the beam beam counter (BBC), which only counts
a fraction of the events, missing in particular low multiplicity events. 

\begin{figure}
\begin{center}\includegraphics[  scale=0.4,
  angle=270,
  origin=c]{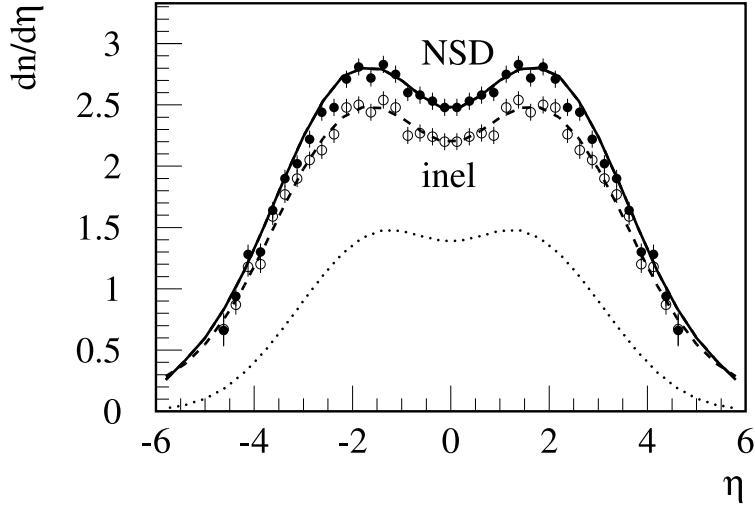}\end{center}
\vspace{-2cm}

\caption{Pseudorapidity distribution for inelastic and NSD events in $pp$
collisions.\label{ua5}The lines are EPOS results, the points data
\cite{ua5}. The dotted lined represents the {}``inner contribution''
to the inelastic distribution (many particles are coming from remnants!).}
\end{figure}

UA5 \cite{ua5} actually used a similar trigger to define non single
diffractive (NSD) events. The NSD differential yield is given as\begin{equation}
\frac{d^{3}n^{\mathrm{NSD}}}{dyd^{2}p_{t}}=\frac{1}{\sigma _{\mathrm{NSD}}}\, \frac{d^{3}\sigma ^{\mathrm{NSD}}}{dyd^{2}p_{t}},\end{equation}
where $\sigma \approx 35$ mb is the NSD $pp$ cross section, and
$d^{3}\sigma ^{\mathrm{NSD}}/dyd^{2}p_{t}$ represents the inclusive
differential cross section for NSD events. To be clear: NSD is not
an absolute definition, it is defined via the acceptance of the UA5
detector! In figure \ref{ua5}, we show the corresponding pseudorapidity
distribution for NSD events, slightly higher than the one for inelastic
events. For the simulation of NSD events, we use simply the same requirement
as used in the experiment (coincidence of charged particles in a forward
and a backward pseudorapidity interval). 

\begin{figure}
\begin{center}\includegraphics[  scale=0.4,
  angle=270,
  origin=c]{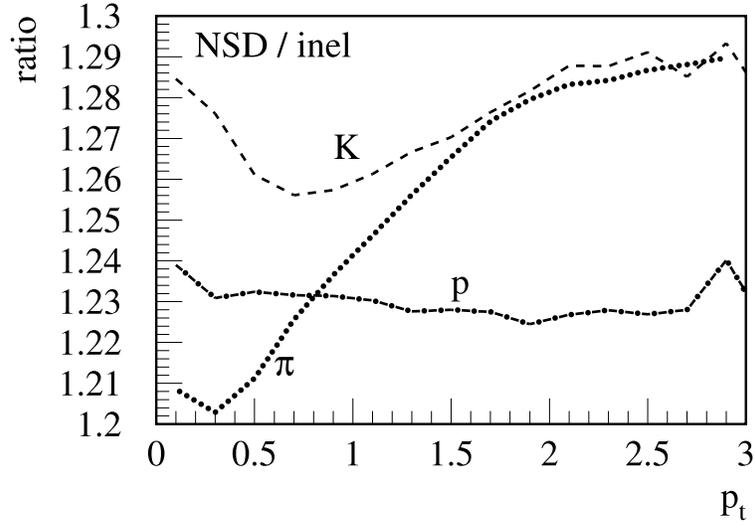}\end{center}
\vspace{-2cm}

\caption{Ratio of the NSD$^{\mathrm{BBC}}$ differential yield to the inelastic
differential yield, in $pp$ collisions, for a pions ($\pi $), kaons
(K), and protons (p).\label{nsdbbc}}
\end{figure}

In case of STAR, one could as well define NSD as the events accepted
by the BBC. Let us do so for the moment, and use the term NSD$^{\mathrm{BBC}}$.
What is actually done is somewhat different. The differential cross
section is multiplied by 30/26, in order to correspond to what Pythia
defines to be non single diffractive, corresponding to 30mb. Let us
call this NSD$^{\mathrm{PYT}}$. Actually the inelastic differential
yield does not change, however, it is interpreted as spectrum for
NSD$^{\mathrm{PYT}}$ events. Then again based on Pythia, it is argued
that the inelastic differential yield for inelastic events is obtained
essentially (with a small correction at small $p_{t}$) by multiplying
with 30/42 (just the ratio of the cross sections), since SD events
do not contribute to particle production. So after all, the originally
measured differential yield (referring to NSD$^{\mathrm{BBC}}$) and
the inelastic one differ essentially by a factor 42/30 = 1.4 -- what
is not at all what we observe, simulating NSD events with the BBC
trigger condition, and comparing with inelastic events. As seen in
fig. \ref{nsdbbc}, the ratio of the NSD$^{\mathrm{BBC}}$ differential
yield to the inelastic differential yield, rather than being 1.4,
differs considerably as a function of $p_{t}$, and in addition depends
on the particle species.

Based on the above discussion, we will simulate NSD$^{\mathrm{BBC}}$
differential spectra (we actually get exactly 26 mb for the NSD$^{\mathrm{BBC}}$
cross section), and compare with STAR's published NSD results, since
they are identical to the NSD$^{\mathrm{BBC}}$ differential spectra.
In this way we avoid all these problems related to Pythia correction
procedures. In the following, NSD refers always to NSD$^{\mathrm{BBC}}$.

\begin{figure}
\begin{center}\includegraphics[  scale=0.4,
  angle=270,
  origin=c]{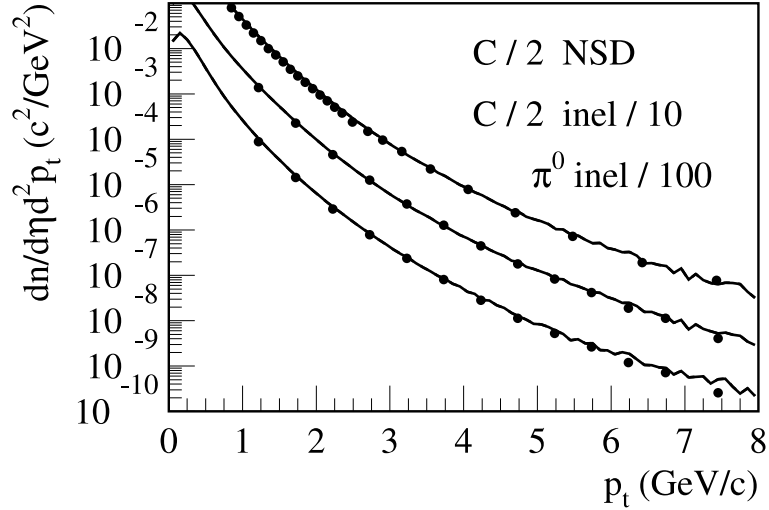}\end{center}
\vspace{-2cm}

\caption{Differential yields in $pp$ collisions as a function of $p_{t}$,
for (from top to bottom) charged particles (over 2) for NSD events,
charged particles (over 2) for inelastic events, and neutral pions
for inelastic events. Lines are EPOS simulations, points are data
from STAR \cite{star-pp} and PHENIX \cite{phenix-a,phenix-pp-piz}.
The two agree within 15\% (over 6 orders of magnitude). \label{ptspectra}}
\end{figure}

In fig. \ref{ptspectra}, we show $p_{t}$ spectra for NSD events,
compared to STAR data \cite{star-pp}, and inelastic events, compared
to PHENIX data \cite{phenix-a,phenix-pp-piz}. Simulation and data
agree within 15\% (over 6 orders of magnitude). This good agreement
is only possible after our re-interpretation of NSD, see the discussion
above. 

\begin{figure}
\begin{center}\includegraphics[  scale=0.4,
  angle=270,
  origin=c]{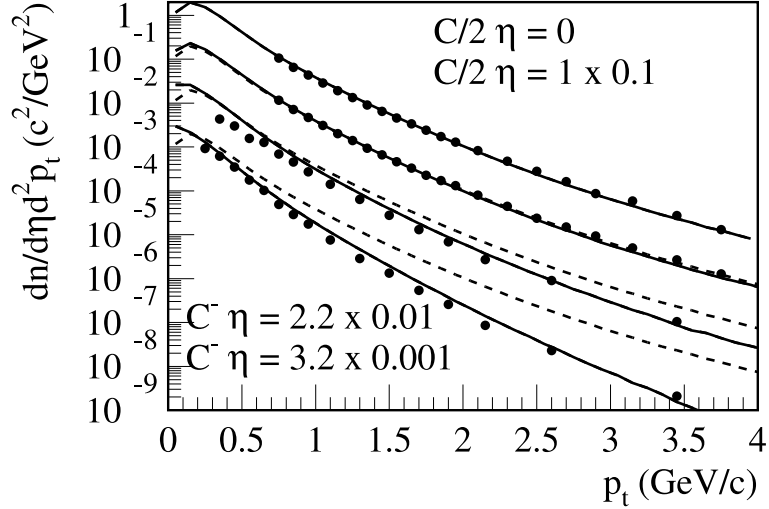}\end{center}
\vspace{-2cm}

\caption{Inelastic differential yields in $pp$ collisions as a function of
$p_{t}$, for (from top to bottom) charged particles (over 2) at $\eta =0$,
at $\eta =1$, negative particles at $\eta =2.2$, at $\eta =3.2$
(always displaced by factors of 10). Lines are EPOS simulations, poins
are data \cite{brahms-b}. We also plot (dashed) the simulation curve
at $\eta =0$, multiplied by 0.1, 0.01, 0.001, to have a reference.\label{ptspectraeta}}
\end{figure}

When studying (later) $dAu$ collisions, there will be plenty of discussion
concerning the (pseudo)rapidity dependence of certain effects. It
is therefore necessary to first check the (pseudo)rapidity dependence
of $p_{t}$ spectra for $pp$. In fig. \ref{ptspectraeta},%
we plot inelastic differential yields as a function of $p_{t}$, at
different pseudorapidities; $\eta =0$, $\eta =1$, $\eta =2.2$,
and $\eta =3.2$ We show EPOS simulations compared to BRAHMS data
\cite{brahms-b}. We also plot (dashed) the simulation curve at $\eta =0$,
multiplied by 0.1, 0.01, 0.001, to have a reference for the results
at the other pseudorapidities. The spectra clearly get softer with
increasing $\eta $.

\section{Results Deuteron-Gold}

\begin{figure}
\begin{center}\includegraphics[  scale=0.4,
  angle=270,
  origin=c]{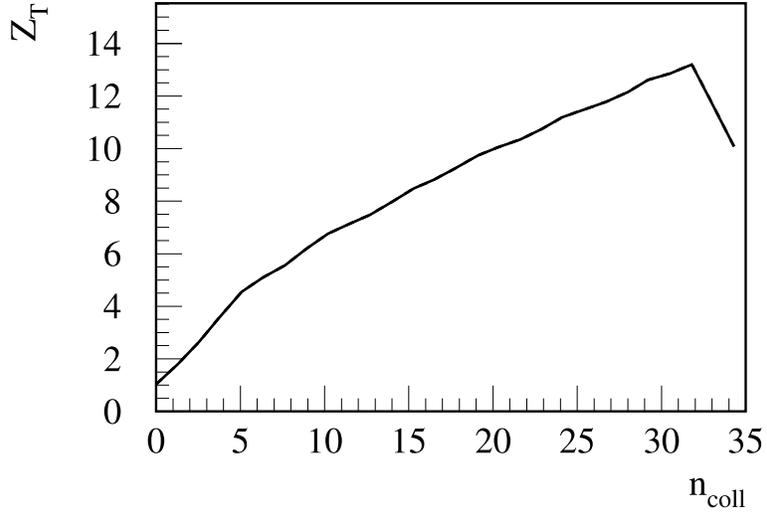}\end{center}
\vspace{-2cm}

\caption{The target $Z$ as a function of centrality, expressed in terms of
the number of binary collisions, for $dAu$. \label{zcental}}
\end{figure}

\begin{figure}
\begin{center}\includegraphics[  scale=0.4,
  angle=270,
  origin=c]{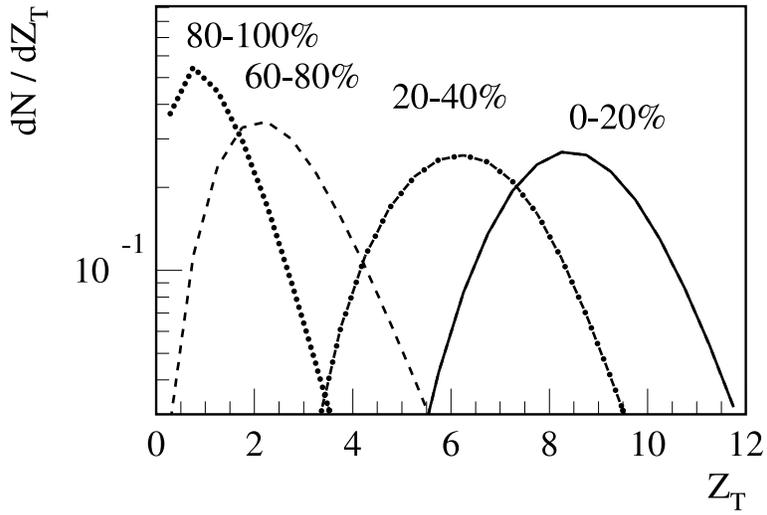}\end{center}
\vspace{-2cm}

\caption{The $Z$ distribution, for different centrality classes. \label{zdistr}}
\end{figure}

All screening effect are linear in $Z$, so it is worthwhile to first
investigate $Z$. In very asymmetric collisions as $dAu$, the projectile
$Z$ is essentially zero, whereas the target $Z$ differers considerably
from zero. As show in fig. \ref{zcental} (and obvious from the definition)
$Z_{T}$ increases linearly with the number of collisions. So $Z$
is essentially a centrality measure. %
In fig. \ref{zdistr}, we show the $Z$ distribution for the different
centrality classes.%
In this way one understands easily how the different centrality classed
are affected by the splitting effects.

\begin{figure}
\begin{center}\includegraphics[  scale=0.4,
  angle=270,
  origin=c]{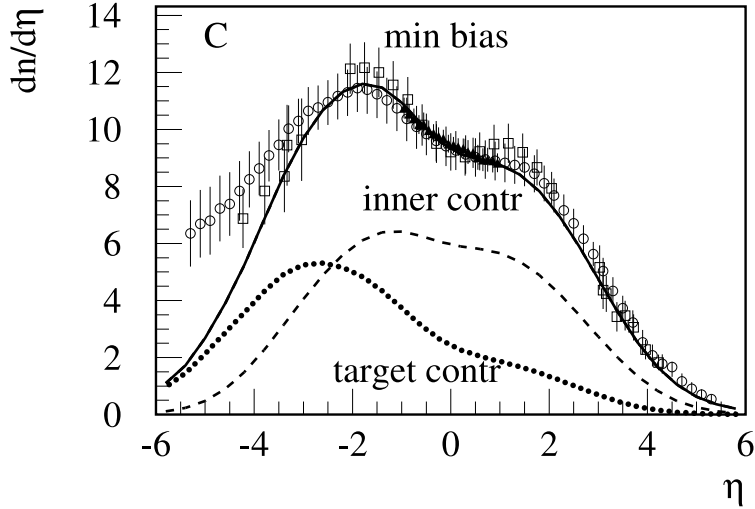}\end{center}
\vspace{-2cm}

\caption{Pseudorapidity spectra in minimum bias $dAu$ collisions. Lines are
EPOS simulations, points are data from PHOBOS \cite{phobos-f} (circles),
STAR \cite{star-b} (triangles), BRAHMS \cite{brahms-c} (squares).
We also show the inner and the outer target contributions to the simulated
distribution.\label{pseudomb}}
\end{figure}

\begin{figure}
\begin{center}\includegraphics[  scale=0.4,
  angle=270,
  origin=c]{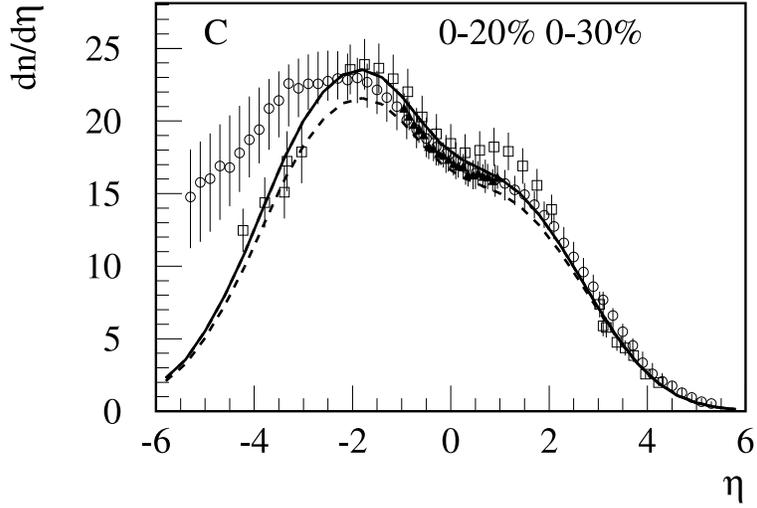}\end{center}
\vspace{-2cm}

\caption{Pseudorapidity spectra for central $dAu$ collisions. Solid lines
are EPOS simulations for 0-20\%, dashed lines are simulations for
0-30\%. Points are data from PHOBOS \cite{phobos-d} (circles), STAR
\cite{star-b} (triangles), BRAHMS \cite{brahms-c} (squares). \label{pseudocentr}}
\end{figure}

\begin{figure}
\begin{center}\includegraphics[  scale=0.4,
  angle=270,
  origin=c]{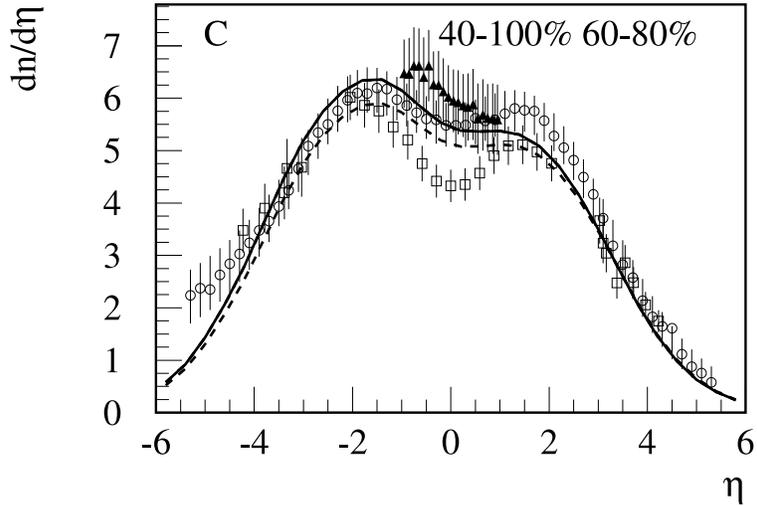}\end{center}
\vspace{-2cm}

\caption{Pseudorapidity spectra for peripheral $dAu$ collisions. Solid lines
are EPOS simulations for 40-100\%, dashed lines are simulations for
60-80\%. Points are data from PHOBOS \cite{phobos-d} (circles), STAR
\cite{star-b} (triangles), BRAHMS \cite{brahms-c} (squares). \label{pseudocentr2}}
\end{figure}

Although we are mainly interested here in transverse momentum spectra,
we still show first of all the pseudorapidity spectra, which finally
determine the normalization of the $p_{t}$ spectra. In fig. \ref{pseudomb},
we show pseudorapidity spectra in minimum bias $dAu$ collisions:
EPOS simulations, compared to data from from PHOBOS \cite{phobos-f},
STAR \cite{star-b}, and BRAHMS \cite{brahms-c}. We also show different
contributions to the simulated distribution. We distinguish inner
and outer (projectile and target) contributions, where the outer contributions
are meant to contain the multiple ladders, originating from ladder
splittings, treated in a {}``collective'' way, as discussed above.
The inner contribution comes from ordinary ladders in the middle.
The asymmetry of the distribution is clearly due to to the target
contribution (the projectile contribution, not shown, is very small).
Also not shown here is the result for {}``no splitting''; for central
collisions, ladder splitting leads to an overall reduction of $dn/d\eta $
of about 30\%. This is due to the fact that first of all the two splitting
effects {}``screening'' and {}``string break delay'' are relatively
small, and secondly work in opposite direction. In figs. \ref{pseudocentr}
and \ref{pseudocentr2}, we show pseudorapidity spectra for central
and peripheral $dAu$ collisions. 


Let us now turn to $p_{t}$ spectra. One of the first observations
concerning $p_{t}$ spectra in $dAu$ collisions was the fact that
not only the nuclear modification factor shows a non-trivial behavior,
but this behavior seems also to be strongly pseudorapidity dependent,
even when varying $\eta $ only by one unit. We want to investigate
this question in the following. 

\begin{figure}
\begin{center}\hspace*{-0.6cm}\includegraphics[  scale=0.35,
  angle=270,
  origin=c]{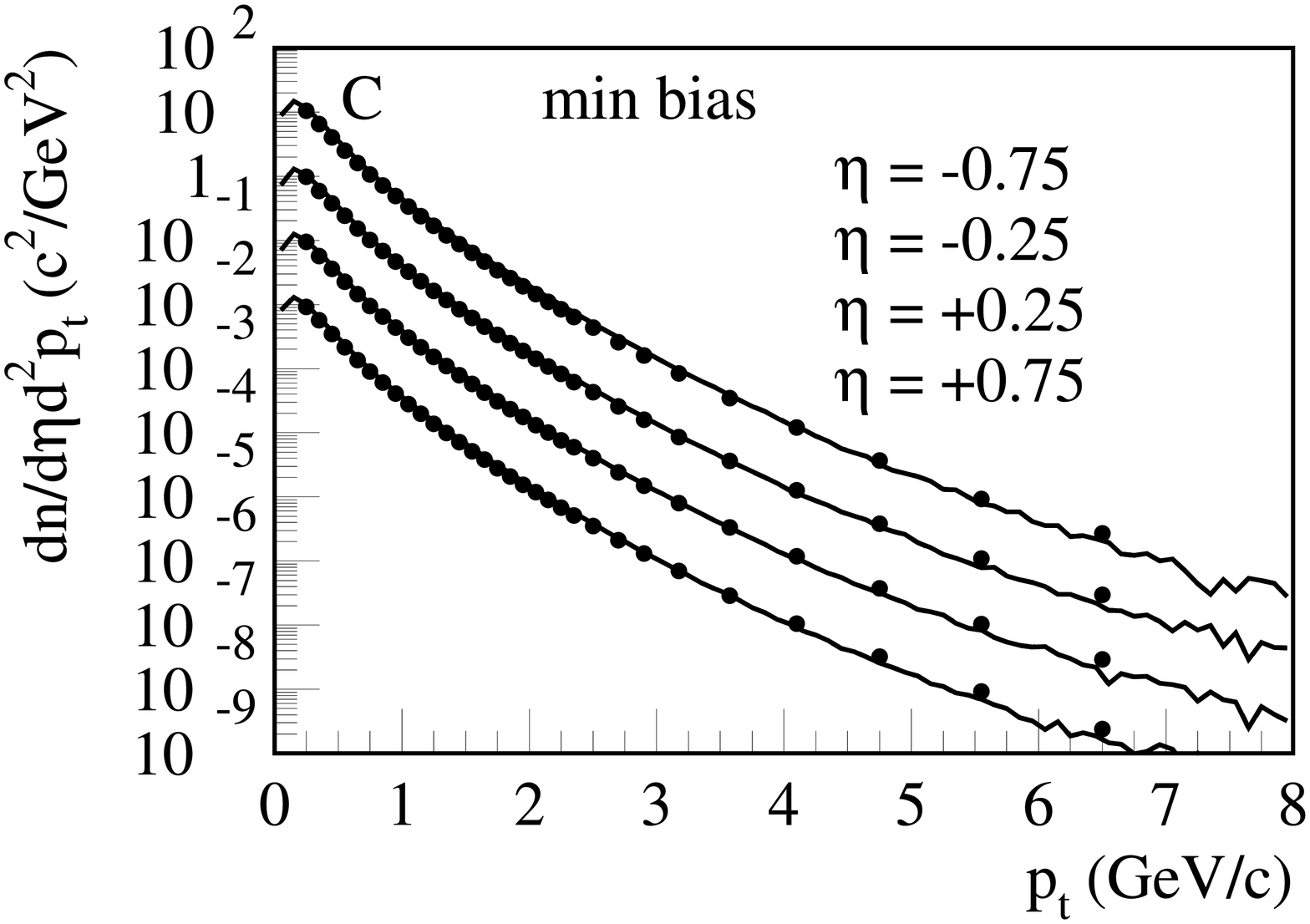}\includegraphics[  scale=0.35,
  angle=270,
  origin=c]{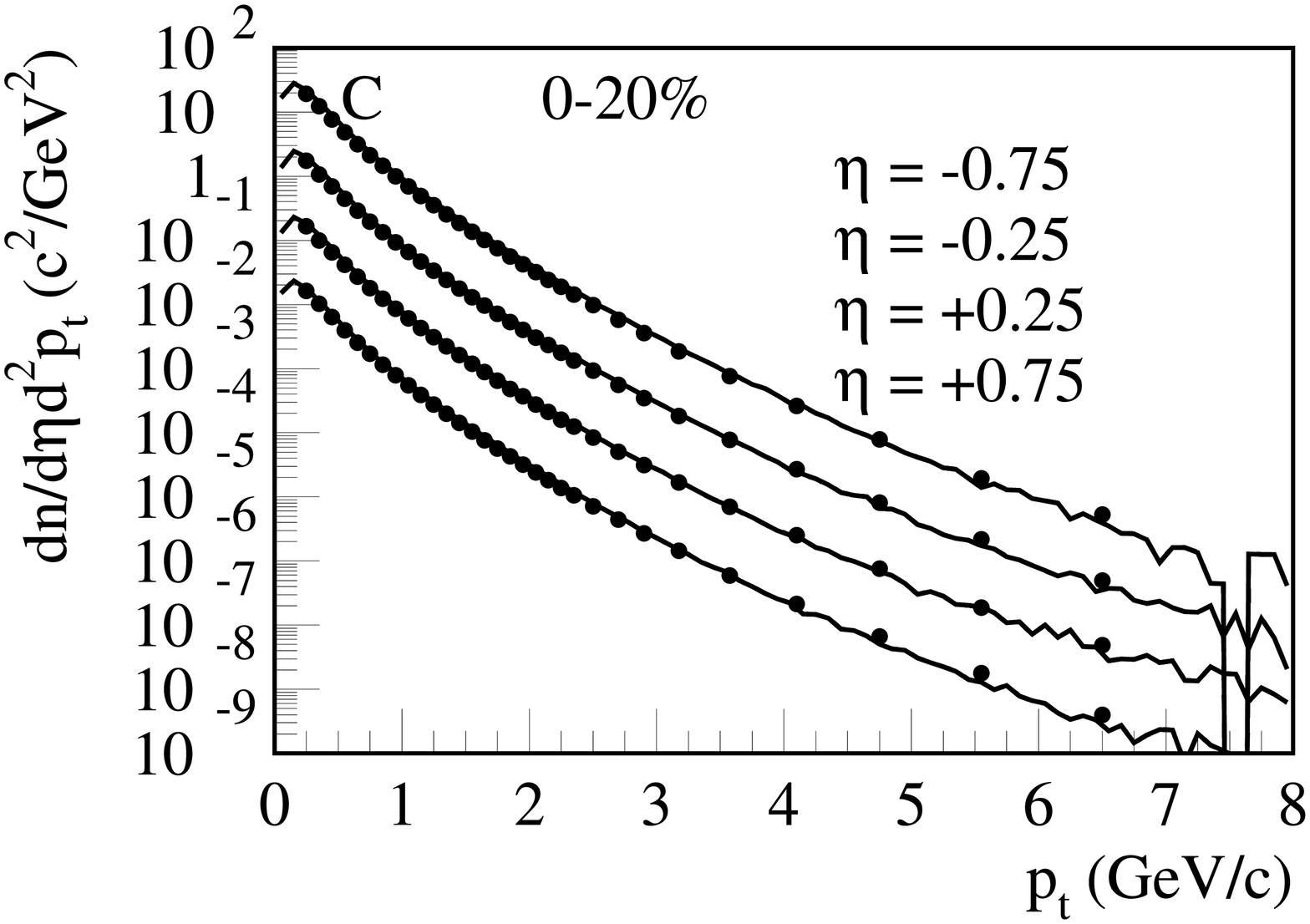}\end{center}

\hspace*{-0.6cm}\includegraphics[  scale=0.35,
  angle=270,
  origin=c]{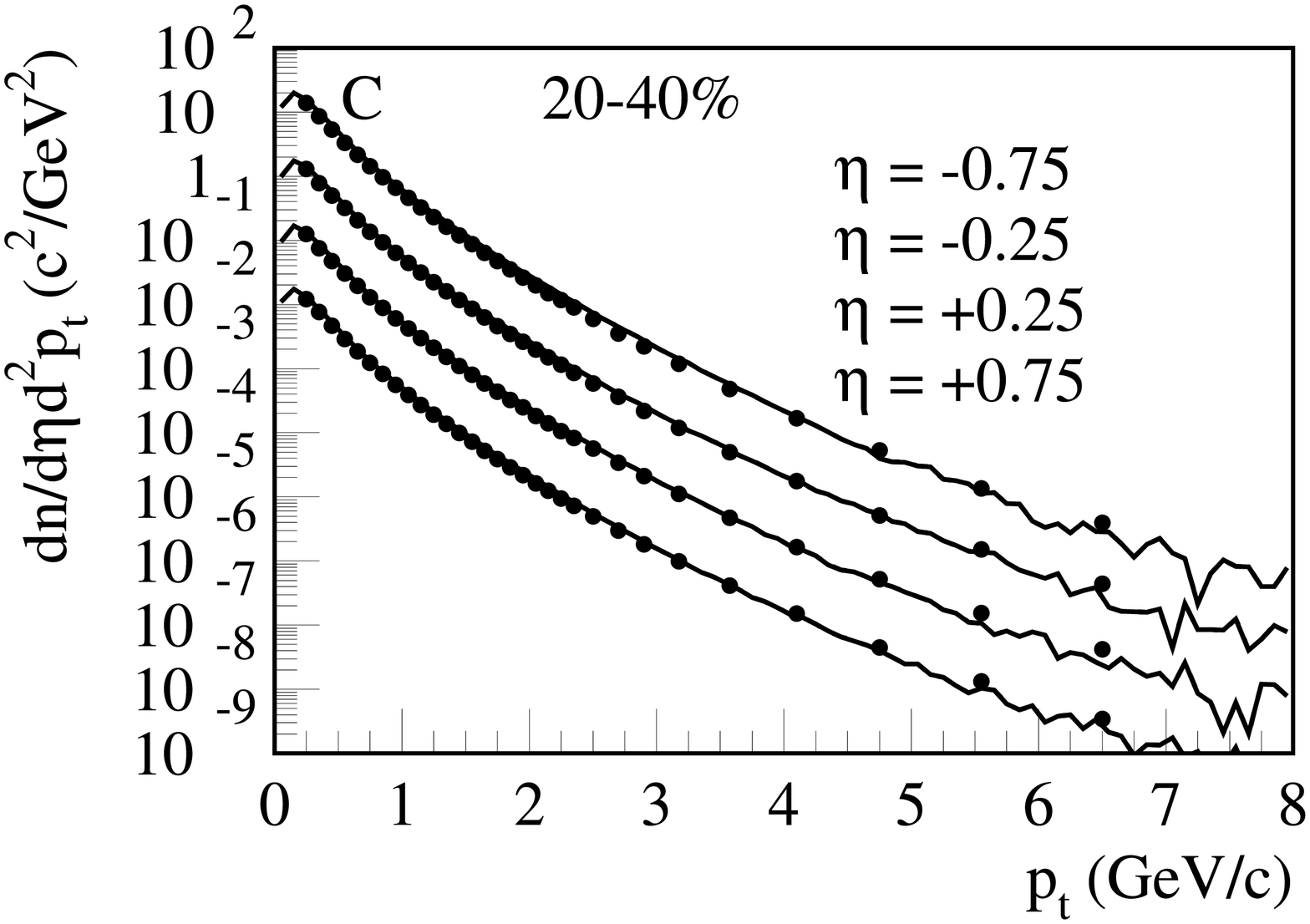}\includegraphics[  scale=0.35,
  angle=270,
  origin=c]{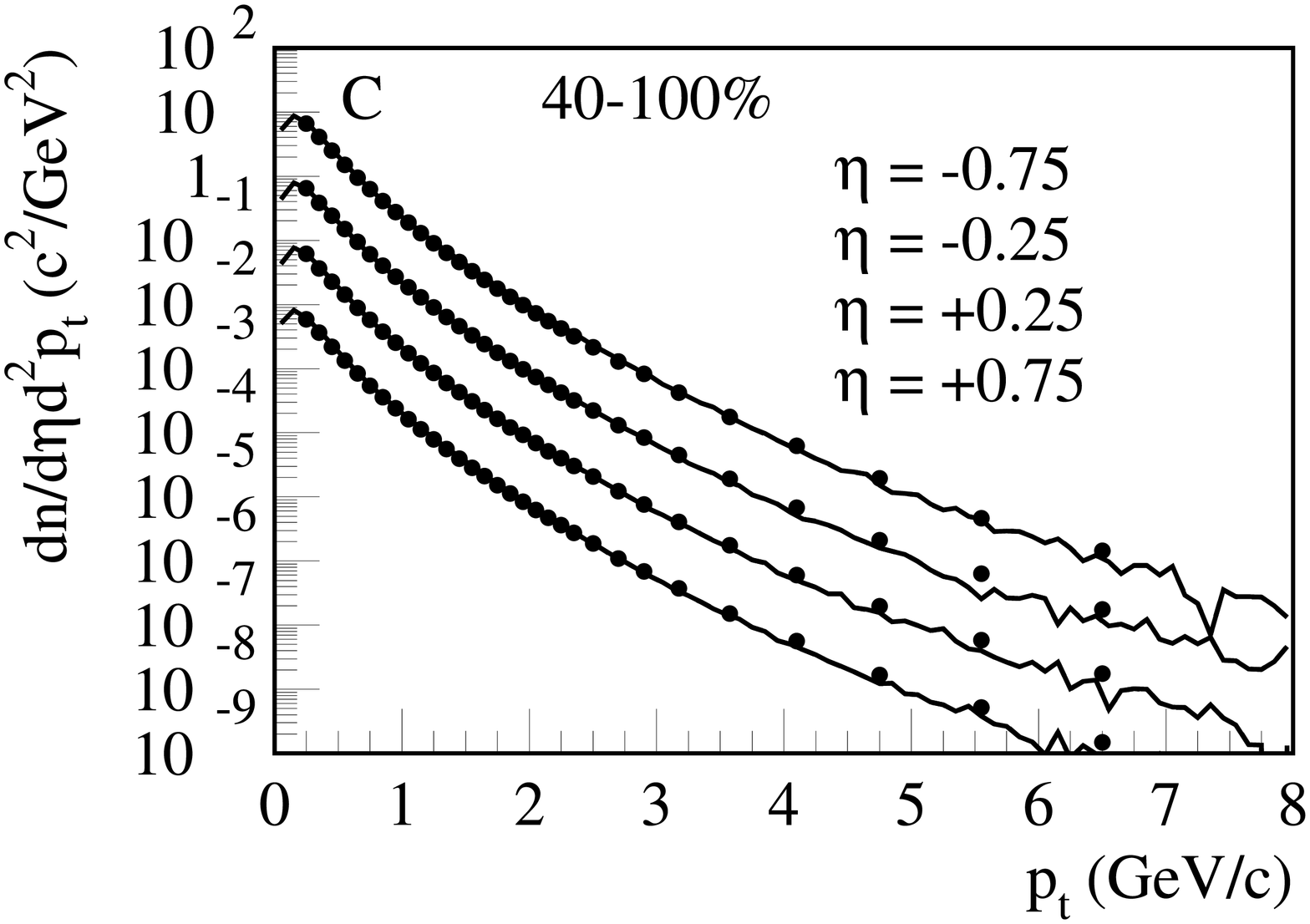}

\caption{Transverse momentum spectra of charged particles in $dAu$ collisions
at different centralities and at different pseudorapidities. The four
figures represent minimum bias, central ($0-20\%$), mid-central ($20-40\%$)
, and peripheral ($40-100\%$) collisions. For each figure, from top
to bottom: $\eta =-0.75$, $\eta =-0.25$, $\eta =0.25$, $\eta =0.75$.
Solid lines are EPOS simulations, points are data \cite{star-b}.
The different curves have been displaced by factors of 10.\label{star-b-1}}
\end{figure}

In fig. \ref{star-b-1}, we show transverse momentum spectra of charged
particles in $dAu$ collisions at different centralities and at different
pseudorapidities. The four figures represent minimum bias, central
($0-20\%$), mid-central ($20-40\%$) , and peripheral ($40-100\%$)
collisions. For each figure, spectra for four pseudorapidity intervals
are shown: $[-1,-0.5]$, $[-0.5,0]$, $[0,0.5]$, $[0.5,1]$. We simply
refer to the corresponding mean values, $\eta =-0.75$, $\eta =-0.25$,
$\eta =0.25$, $\eta =0.75$. For better visibility, the different
curves have been displaced by factors of 10. Solid lines are EPOS
simulations, points are data \cite{star-b}, both agree within 10-20\%.
Although looking directly at spectra does not really allow to see
systematic differences between the different curves, it is still useful
to first check that the absolute curves agree, before investigating
ratios. 

\begin{figure}
\begin{center}\hspace*{-0.6cm}\includegraphics[  scale=0.35,
  angle=270,
  origin=c]{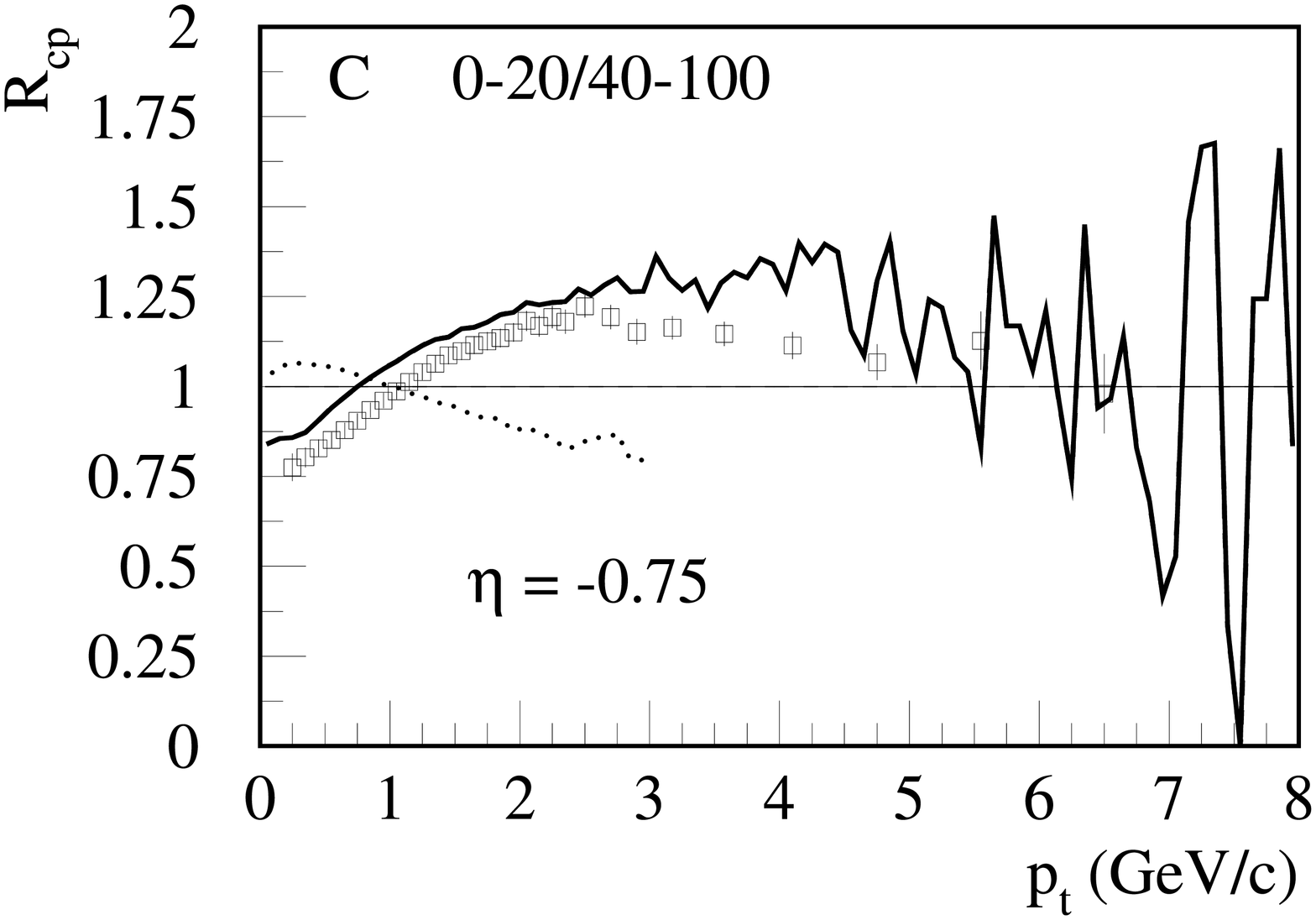}\includegraphics[  scale=0.35,
  angle=270,
  origin=c]{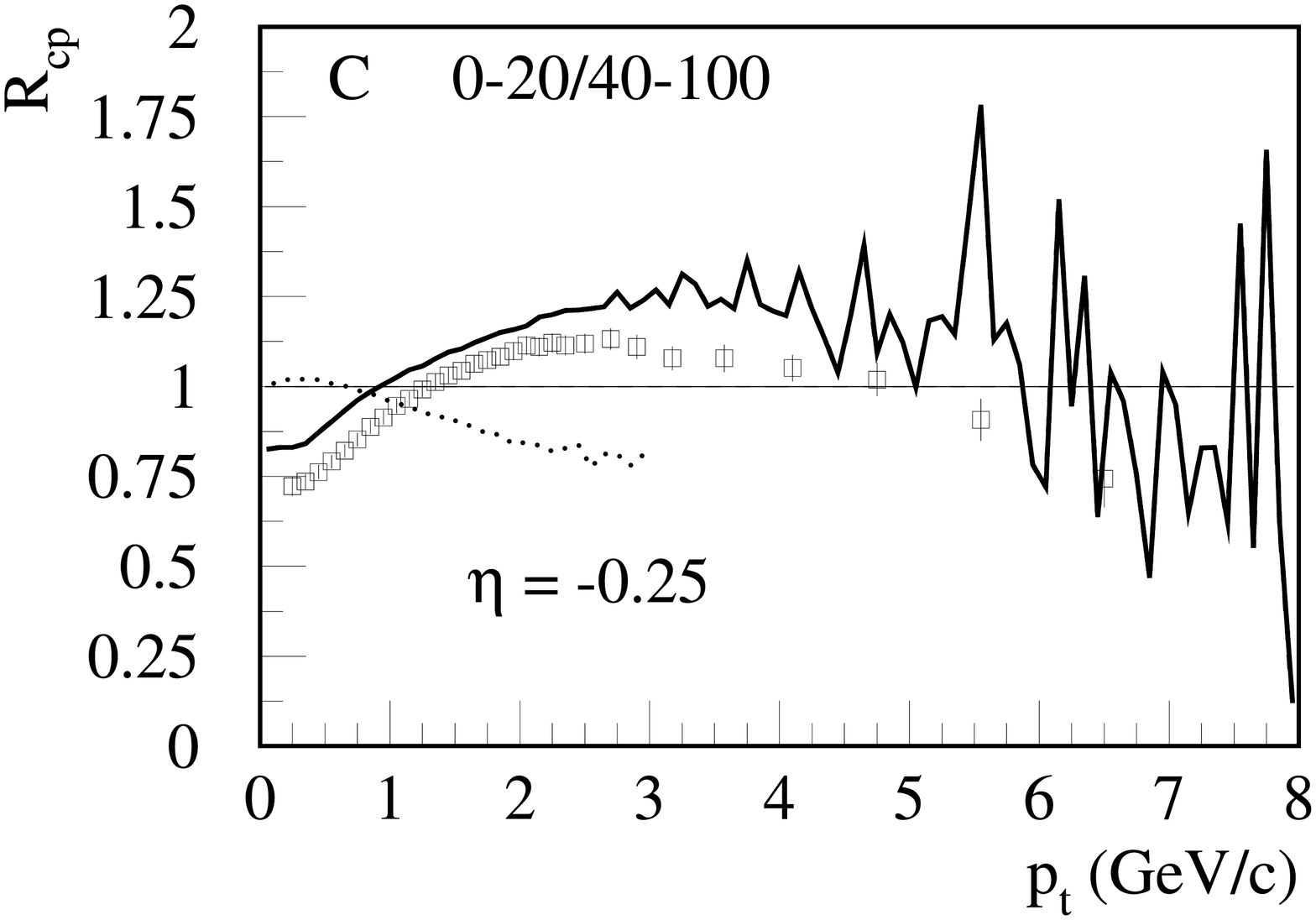}\end{center}

\hspace*{-0.6cm}\includegraphics[  scale=0.35,
  angle=270,
  origin=c]{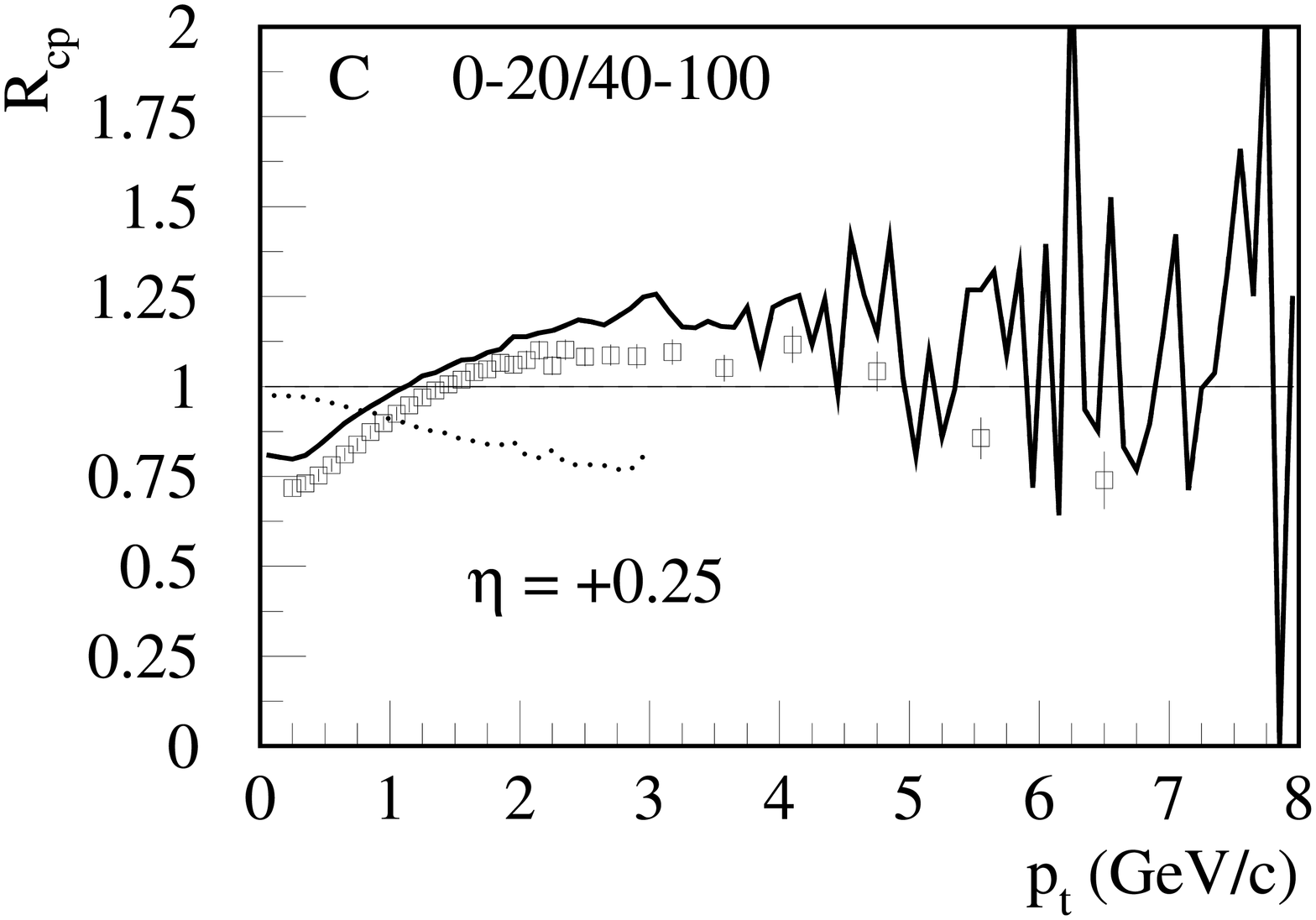}\includegraphics[  scale=0.35,
  angle=270,
  origin=c]{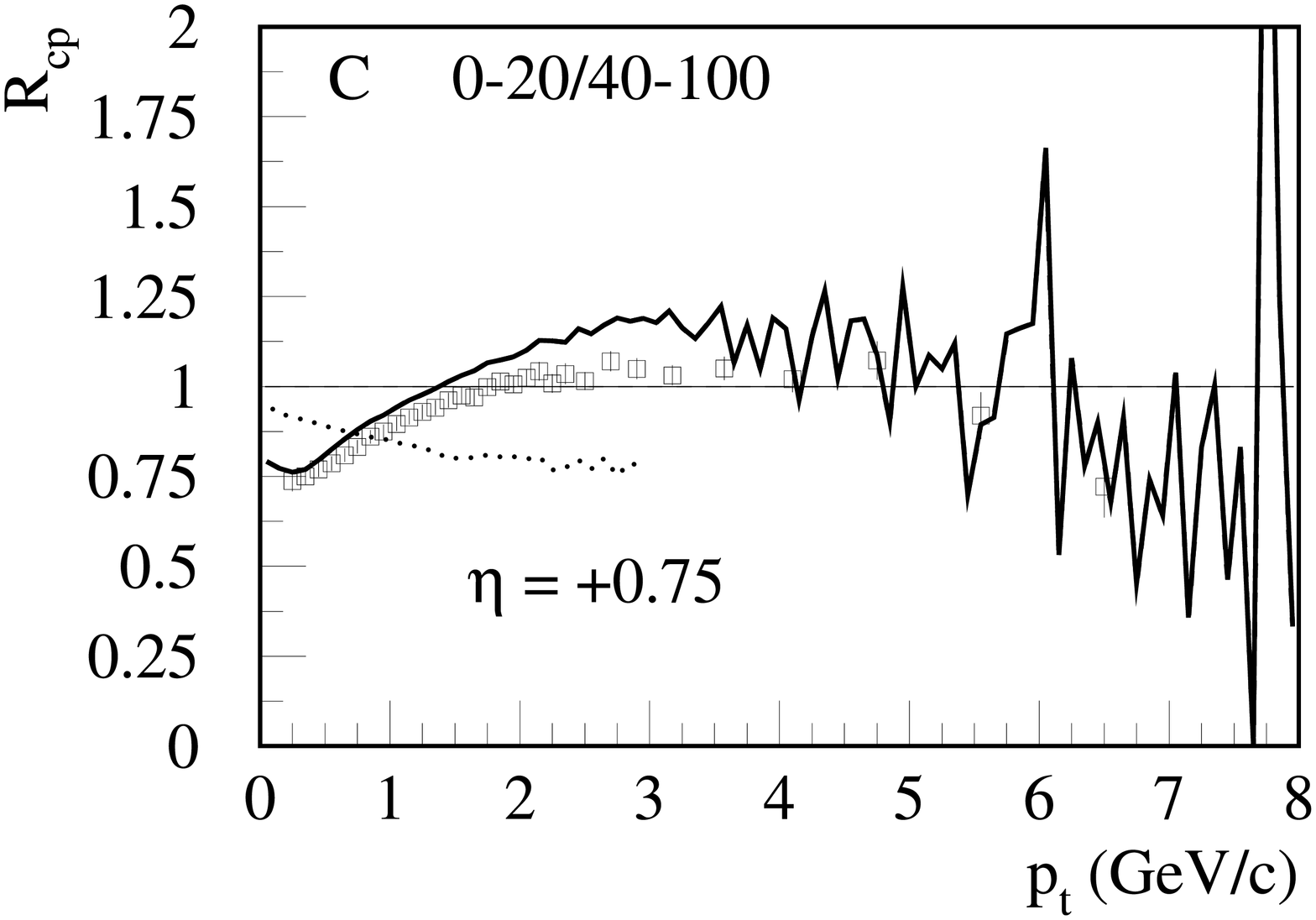}

\caption{$R_{cp}$ ratios at different pseudorapidities ($\eta =-0.75$, $\eta =-0.25$,
$\eta =0.25$, $\eta =0.75$). Solid lines are EPOS simulations, points
are data \cite{star-b}. The dotted lines are EPOS simulations, with
parton ladder splitting turned off. \label{star-b-2}}
\end{figure}

To observe any {}``anomalous'' behavior, one usually plots ratios,
like the nuclear modification factor, defined earlier. The disadvantage
is the fact that the corresponding $pp$ spectrum has to be known,
with a sufficient precision. An alternative procedure is the use of
ratios of central to peripheral results,\begin{equation}
R_{cp}=\frac{1}{N_{\mathrm{coll}}^{\mathrm{central}}}\, \frac{dn^{\mathrm{central}}}{d^{2}p_{t}\, dy}\, /\, \frac{1}{N_{\mathrm{coll}}^{\mathrm{peripheral}}}\, \frac{dn^{\mathrm{peripheral}}}{d^{2}p_{t}\, dy}\, .\end{equation}
 In fig. \ref{star-b-2}, we show the $R_{cp}$ ratios at different
pseudorapidities ($\eta =-0.75$, $\eta =-0.25$, $\eta =0.25$$\eta =0.75$).
Here, central refers to $0-20\%$ and peripheral to $40-100\%$. Solid
lines are EPOS simulations, points are data \cite{star-b}. We also
show the corresponding EPOS results, with parton ladder splitting
turned off. These curves are cut off at $p_{t}=3$ GeV/c, to avoid
that the strong statistical fluctuations spoil the figure. The no-parton-ladder-splitting
curve increases slowly with $p_{t}$, in the shown range it stays
well below one. It will finally reach one. %
The full EPOS simulations show quite a different behavior, the ratio
$R_{cp}$ increases strongly between 1 and 2 GeV/c, to stay constant
(or decrease) beyond. The statistical fluctuations do not really allow
very precise predictions beyond 4 GeV/c. The strong increase between
1 and 2 GeV/c is due to collective hadronization on the target side,
which leads to an increased transverse momentum production. As can
be seen from fig. \ref{pseudomb}, target side hadronization extends
even to forward pseudorapidities, so it is quite visible in the whole
$\eta $ range {[}-1,1{]}. The effect is simply somewhat stronger
at backward compared to forward rapidity, since target hadronization
contributes more. But the difference is not so big. The increase of
$R_{cp}$ with $p_{t}$ is partly also due to the momentum transfer
from the elastic splitting, which should affect equally backward and
forward pseudorapidities. The variation of the shape of $R_{cp}$
with pseudorapidity is quite small, the main modification is actually
an overall factor due to the fact, that the particle density increases
towards smaller pseudorapidities, as seen from fig. \ref{pseudomb}.

\begin{figure}
\begin{center}\includegraphics[  scale=0.4,
  angle=270,
  origin=c]{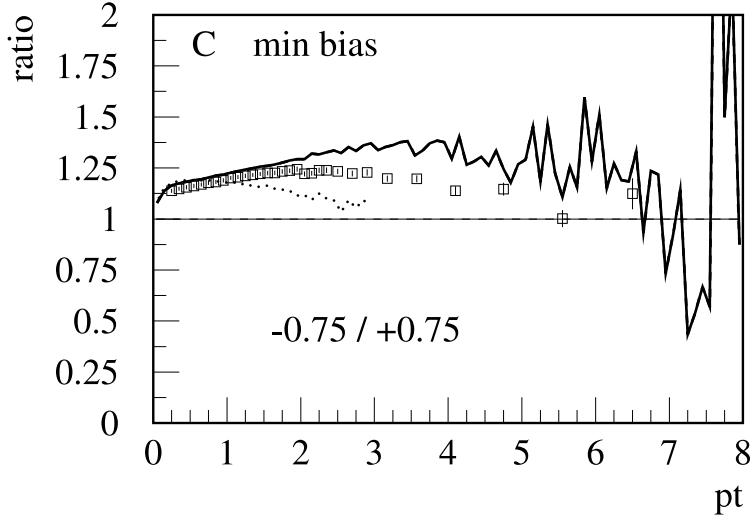}\end{center}
\vspace{-2cm}

\caption{Ratio of charged particle spectra at backward to forward pseudorapidities
($\eta =-0.75$/ $\eta =0.75$), in minimum bias $dAu$ collisions.
Solid lines are EPOS simulations, points are data \cite{star-b}.
The dotted lines are EPOS simulations, with parton ladder splitting
turned off.\label{star-b-3}}
\end{figure}

A direct way to investigate the pseudorapidity dependence of spectra
is provided by the ratio of spectra at backward to forward pseudorapidities,
like $\eta =-0.75$/ $\eta =0.75$, as shown in fig. \ref{star-b-3}.
Here, one observes a slight increase between 0 and 2 GeV/c. This means
that the $R_{cp}$ at backward pseudorapidity increases a bit more
than the one at forward pseudorapidity, which we understand such that
there is somewhat more target side collective hadronization at backward
pseudorapidity.

\begin{figure}
\begin{center}\includegraphics[  scale=0.4,
  angle=270,
  origin=c]{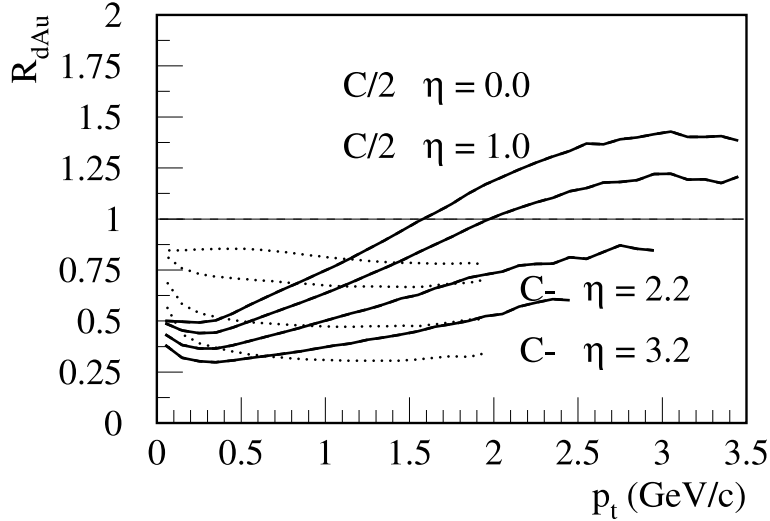}\end{center}
\vspace{-2cm}

\caption{Nuclear modification factors $R_{dAu}$ for charged particles in
minimum bias $dAu$ collisions, at different pseudorapidities: $\eta =0,$
$\eta =1,$ $\eta =2.2,$ $\eta =3.2$ (from top to bottom). The solid
lines are full EPOS simulations, the dotted lines are EPOS with parton
ladder splitting turned off. \label{epospteta}}
\end{figure}

We now consider an even larger pseudorapidity variation: we investigate
how the nuclear modification factors vary in the pseudorapidity range
0 to 3.2. Before comparing to data, we show the results of full EPOS
simulations, as well as those with parton ladder splitting turned
off. In fig. \ref{epospteta}, we show the nuclear modification factors
for charged particles in minimum bias $dAu$ collisions, at different
pseudorapidities: $\eta =0,$ $\eta =1,$ $\eta =2.2,$ $\eta =3.2$.
Whereas the no-splitting curves hardly change with $p_{t}$, and decrease
with pseudorapidity, the full calculations show of course the same
decrease with pseudorapidity, but all curves increase substantially
with $p_{t}$ between 1 and 3 GeV/c. This confirms the observation
already made earlier by studying the variation in the $\eta $ range
-1 to 1.

\begin{figure}
\begin{center}\includegraphics[  scale=0.48,
  angle=270,
  origin=c]{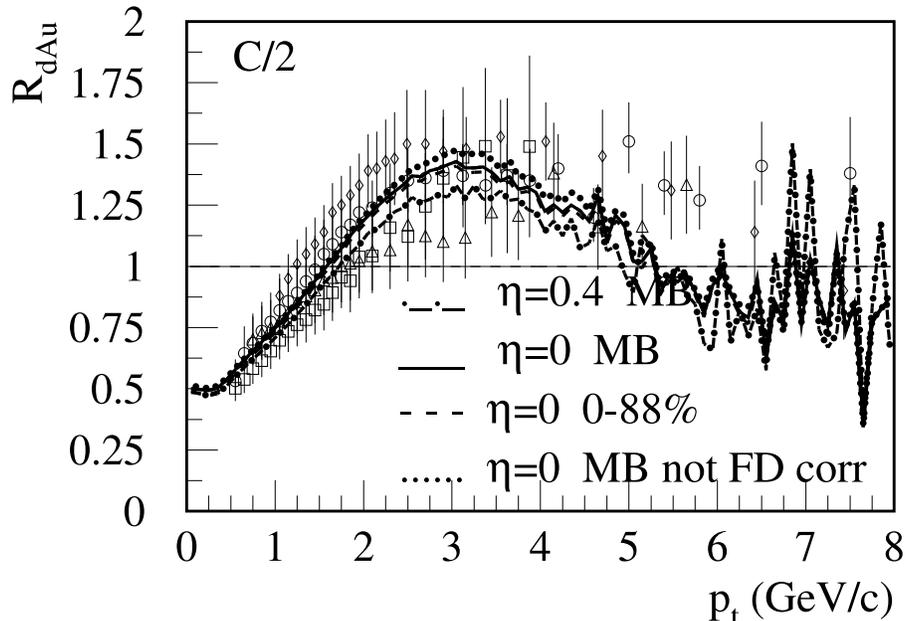}\end{center}
\vspace{-2.8cm}

\caption{Nuclear modification factors $R_{dAu}$ for charged particles in
minimum bias $dAu$ collisions at (or close to) $\eta =0$. The different
lines are full EPOS simulations: minimum bias at $\eta =0$ (full),
at $\eta =0.4$ (dashed-dotted), not feed down corrected (dotted),
0-88\% centrality, at $\eta =0$ (dashed). The points are minimum
bias data at $\eta =0$ from STAR \cite{star-a} (rhombs), at $\eta =0.4$
from PHOBOS \cite{phobos-b} (squares), 0-88\% centrality data at
$\eta =0$ from PHENIX \cite{phenix-a} (circles), minimum bias results
at $\eta =0$ from BRAHMS \cite{brahms-b} (triangles). \label{eta0}}
\end{figure}

In the following we will compare the simulations with data from all
the four RHIC experiments. In fig. \ref{eta0}, we collect all published
data on charged particle nuclear modification factors in minimum bias
$dAu$ collisions at (or close to) $\eta =0$, together with the corresponding
simulations. We show minimum bias results at $\eta =0$ from STAR
\cite{star-a}, at $\eta =0.4$ from PHOBOS \cite{phobos-b}, 0-88\%
centrality results at $\eta =0$ from PHENIX \cite{phenix-a}, and
minimum bias data at $\eta =0$ from BRAHMS \cite{brahms-b}. We also
show minimum bias EPOS simulations at $\eta =0$, at $\eta =0.4$,
not feed down corrected minimum bias simulations, and 0-88\% centrality
results at $\eta =0$.%
We first of all observe that the different simulation results are
quite close to each other, so changing slightly the pseudorapidity,
the centrality definition, doing or not feed down corrections, does
not affect the final result too much. The variation of the experimental
data is much bigger. On the upper end we have the STAR data, but based
on the above discussion on $pp$ results, we expect that the $pp$
reference spectrum is 10-20\% too low, which means $R_{dAu}$ is 10-20\%
too high. The corresponding reduction would bring the STAR data down
to the EPOS simulation curve (full line), and agree with the PHENIX
data, and with the PHOBOS data. BRAHMS is on the lower end, but within
the error bars compatible with the simulation curve. 

\begin{figure}
\begin{center}\includegraphics[  scale=0.48,
  angle=270,
  origin=c]{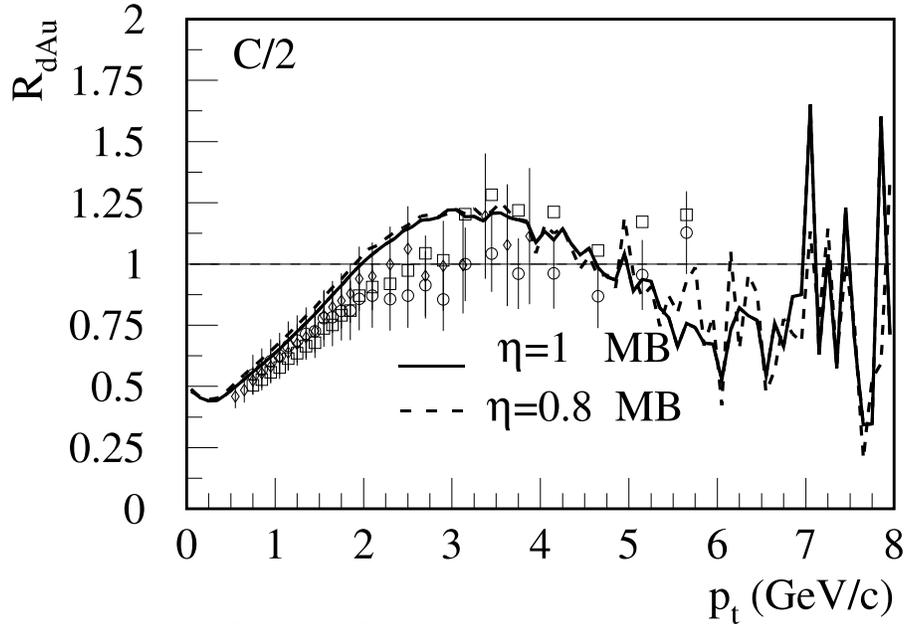}\end{center}
\vspace{-2.8cm}

\caption{Nuclear modification factors $R_{dAu}$ for charged particles in
minimum bias $dAu$ collisions at (or close to) $\eta =1$. The different
lines are full EPOS simulations: minimum bias at $\eta =1$ (full),
at $\eta =0.8$ (dashed). The points are minimum bias data at $\eta =0.8$
from PHOBOS \cite{phobos-b} (rhombs), at $\eta =1$ from BRAHMS \cite{brahms-b}
(circles). The squares represent $R_{dAu}$ calculated from the BRAHMS
$dAu$ $p_{t}$ spectrum and the $pp$ simulation result.\label{eta1}}
\end{figure}

In fig. \ref{eta1}, we consider charged particle nuclear modification
factors in minimum bias $dAu$ collisions at (or close to) $\eta =1$,
together with the corresponding simulations. We show minimum bias
data at $\eta =0.8$ from PHOBOS \cite{phobos-b}, at $\eta =1$ from
BRAHMS \cite{brahms-b}. We also show minimum bias EPOS simulations
at $\eta =1$, at $\eta =0.8$, both being very close to each other.
The data are somewhat lower, but the curves are within the error bars.
A systematic difference may again be due to the $pp$ reference. To
investigate this, we also plot a {}``mixed'' $R_{dAu}$: the nuclear
spectrum is taken from BRAHMS, but we use the EPOS $pp$ reference.
The result (squares) now exceeds the simulation curve.

\begin{figure}
\begin{center}\hspace*{-0.6cm}\includegraphics[  scale=0.35,
  angle=270,
  origin=c]{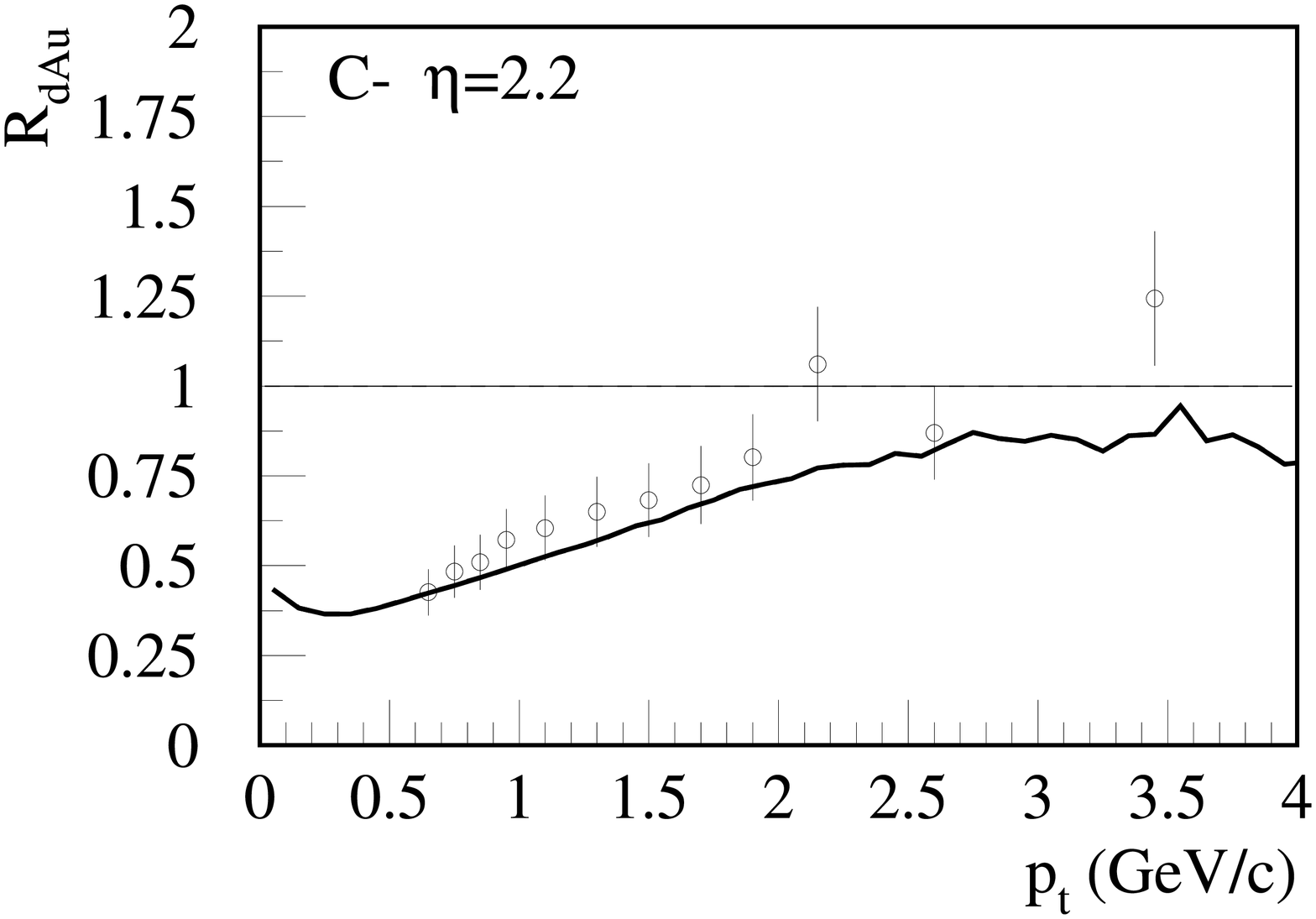}\includegraphics[  scale=0.35,
  angle=270,
  origin=c]{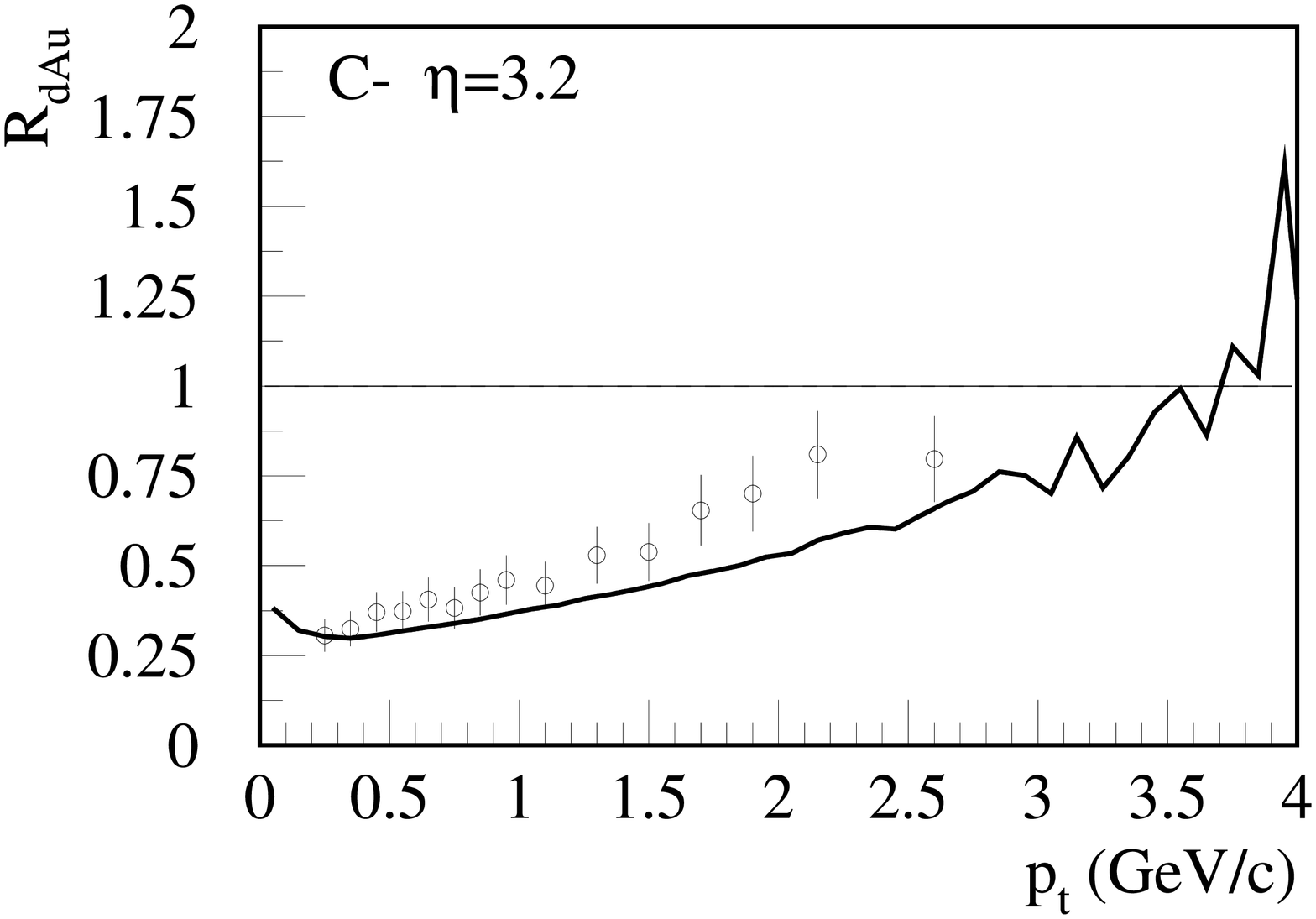}\end{center}

\caption{Nuclear modification factors $R_{dAu}$ for charged particles in
minimum bias $dAu$ collisions at $\eta =2.2$ (left) and $\eta =3.2$
(right). The lines are full EPOS simulations, the points are data
from BRAHMS \cite{brahms-b}.\label{eta2}}
\end{figure}

In fig. \ref{eta2}, we finally compare EPOS simulations and data
from BRAHMS \cite{brahms-b} at $\eta =2.2$ and $\eta =3.2$. Data
and simulations agree quite well.

\section{Summary}

In this paper, we have presented a phenomenological approach, called
EPOS, based on the parton model, but going much beyond. There are
two very important {}``nuclear effects'': elastic and inelastic
parton ladder splitting, which in principle occurs already in $pp$
scattering, but which becomes really visible when systems with large
numbers of partons (like nuclei) are involved.

Elastic splitting is in fact related to screening and saturation,
but much more important is the inelastic contribution, being crucial
to understand the data. 

The main effect (at least concerning the observables investigated
in this paper) is due to the fact that inelastic splitting (bifurcation
of parton ladders), leads to a modified hadronization process, a {}``collective
hadronization'' of multiple, parallel parton ladders, on the target
side, in case of $dAu$. This is the equivalent of string fusion,
if one uses the language of strings. But contrary to the usual string
fusion picture, here we do not have complete ladders which behave
collectively, but only the bifurcated ones on the target side. 

Concerning $p_{t}$ spectra, the main effect of the collective hadronization
is a $p_{t}$ broadening. This is certainly what is needed, but real
evidence for our picture can only come from a very detailed comparison
with all corresponding data currently available. For this purpose
we considered all published nuclear modification factor data, concerning
charged particles, from all the four RHIC experiments. 

We investigated in detail the rapidity dependence of nuclear effects,
which is actually relatively weak in the model, in perfect agreement
with the data, if the latter ones are interpreted correctly, and if
one considers really ALL available data, and not just a convenient
subset.

\appendix

\newpage
\section{Appendix on Multiple Scattering Theory\label{Appendix-Multiple}}

We first consider inelastic proton-proton scattering. We imagine an
arbitrary number of elementary interactions to happen in parallel,
where an interaction may be elastic or inelastic. The inelastic amplitude
is the sum of all such contributions,

\begin{center}\includegraphics[  scale=0.5]{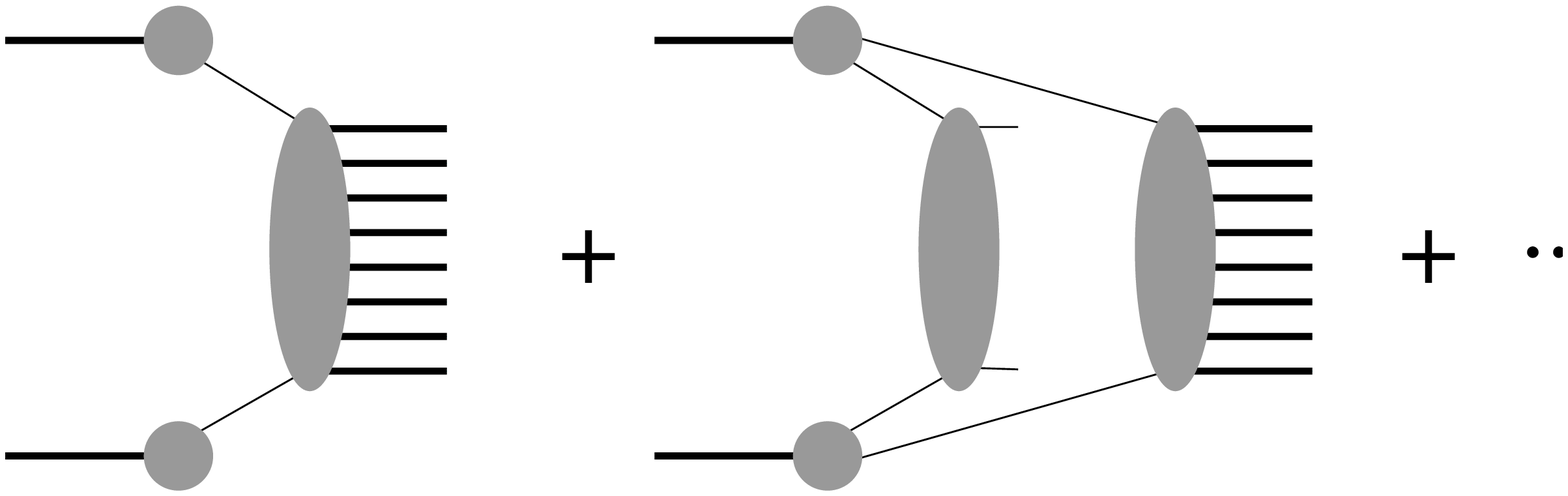}\end{center}

\noindent with at least one inelastic elementary interaction involved.
To calculate cross sections, we need to square the amplitude, which
leads to many interference terms, like for example

\noindent \begin{center}\includegraphics[  scale=0.4]{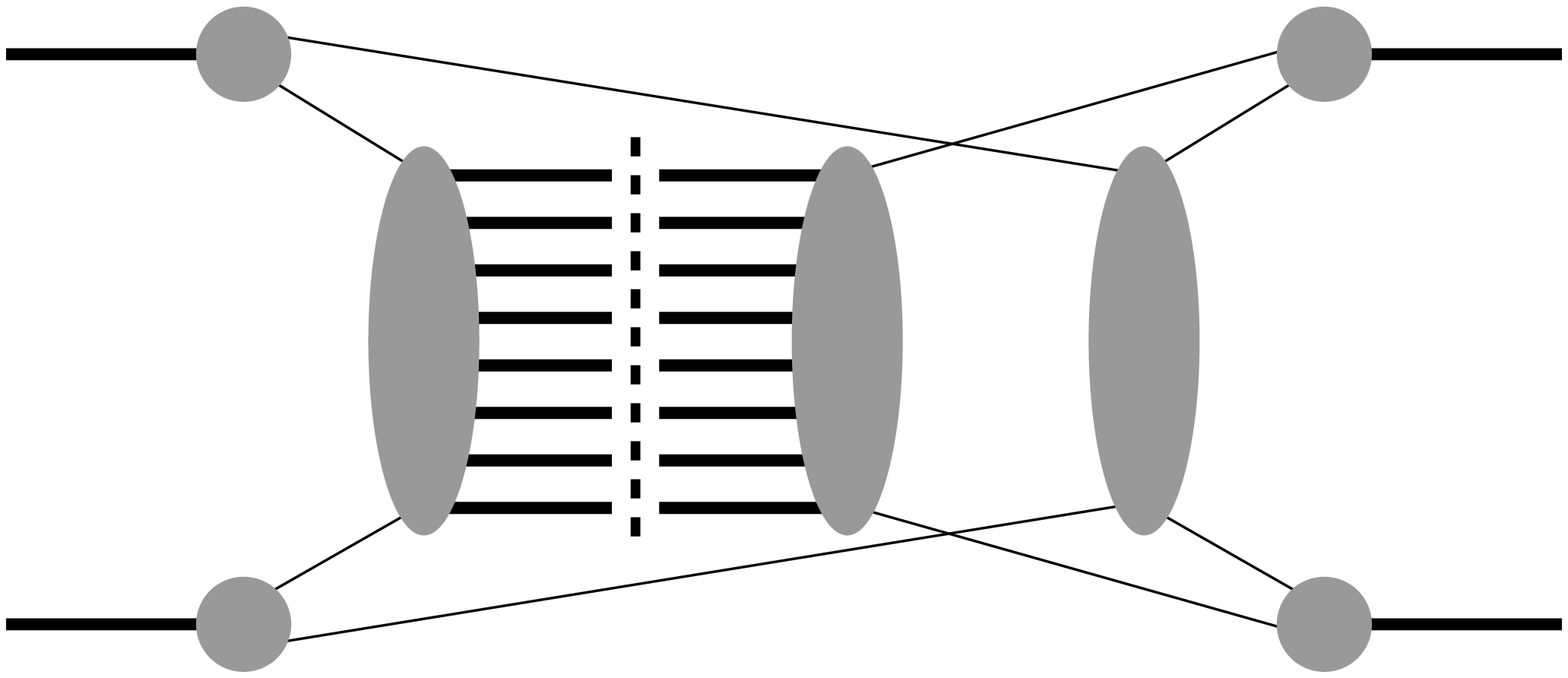}\end{center}

\noindent which represents interference between the first and the
second diagram of the above-mentioned sum of terms. Using the above
notations, we may represent the left part of the diagram as a cut
diagram, conveniently plotted as a dashed line:

\begin{center}\includegraphics[  scale=0.4]{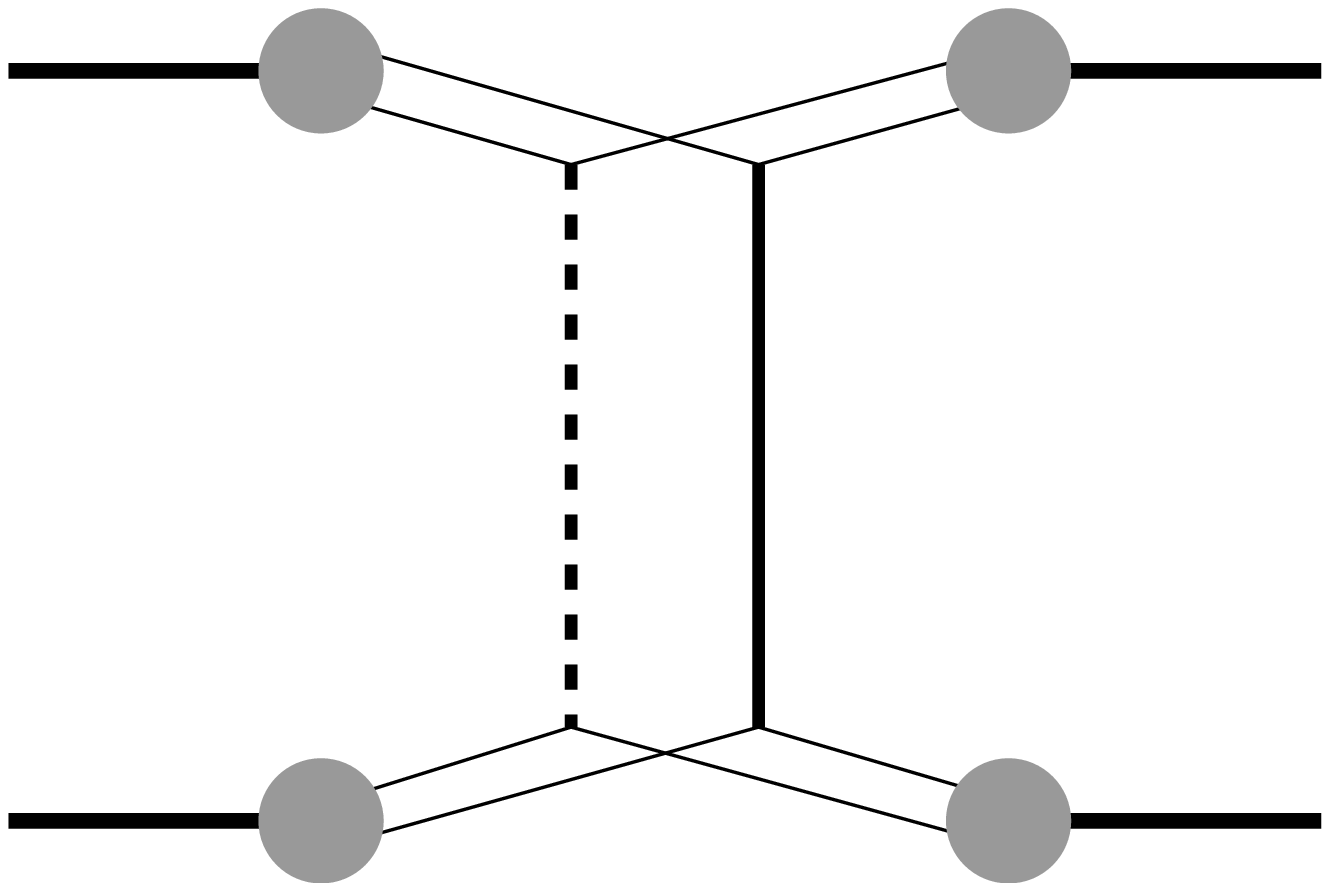}\end{center}

\noindent The squared amplitude is now the sum over many such terms
represented by solid and dashed lines.

\begin{figure}
\begin{center}\includegraphics[  scale=0.35]{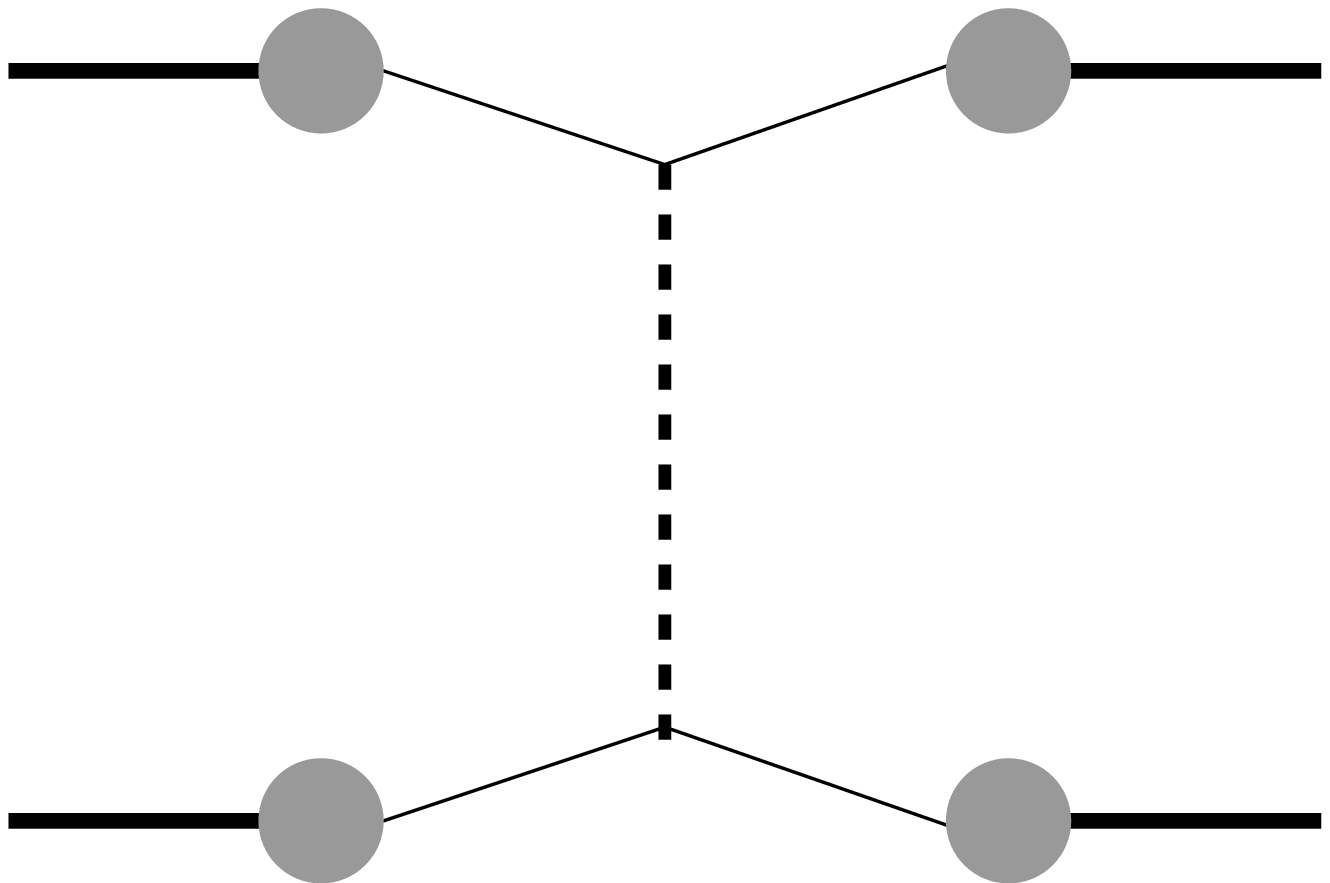}$\qquad $\includegraphics[  scale=0.35]{t7}$\qquad $\includegraphics[  scale=0.35]{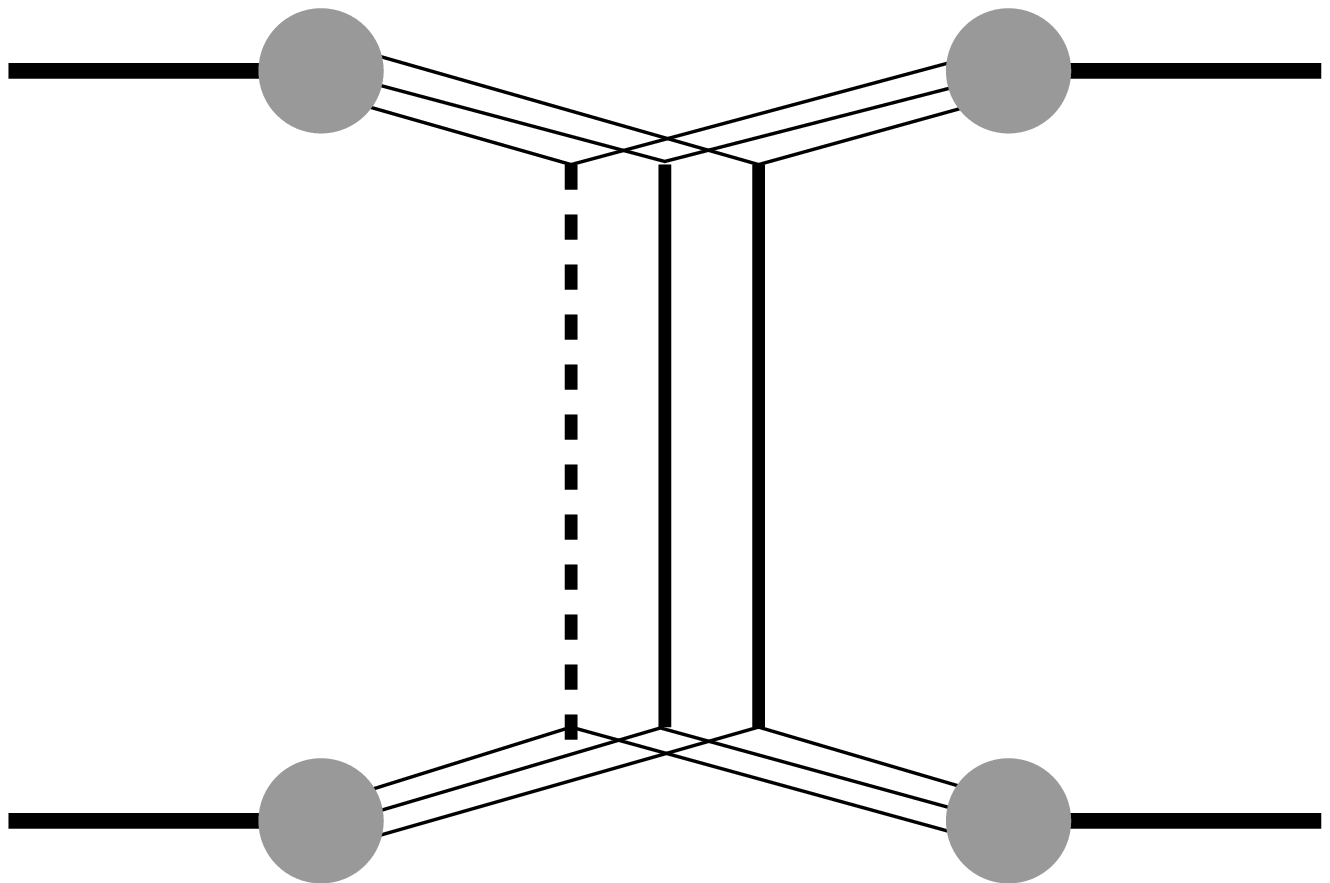}\end{center}

\caption{Class of terms corresponding to one inelastic interaction.\label{t7c}}
\end{figure}

\begin{figure}
\begin{center}\includegraphics[  scale=0.35]{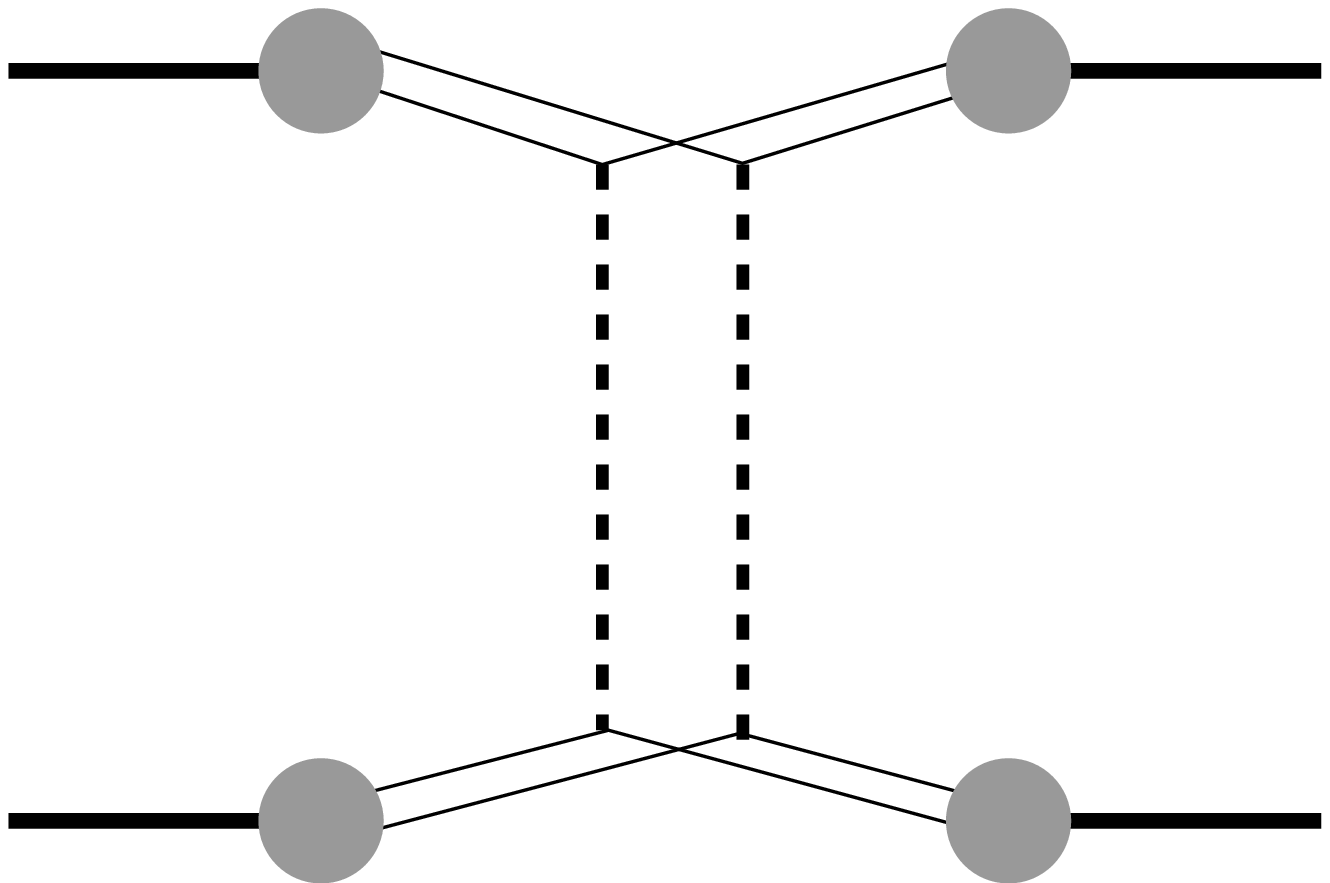}$\qquad $\includegraphics[  scale=0.35]{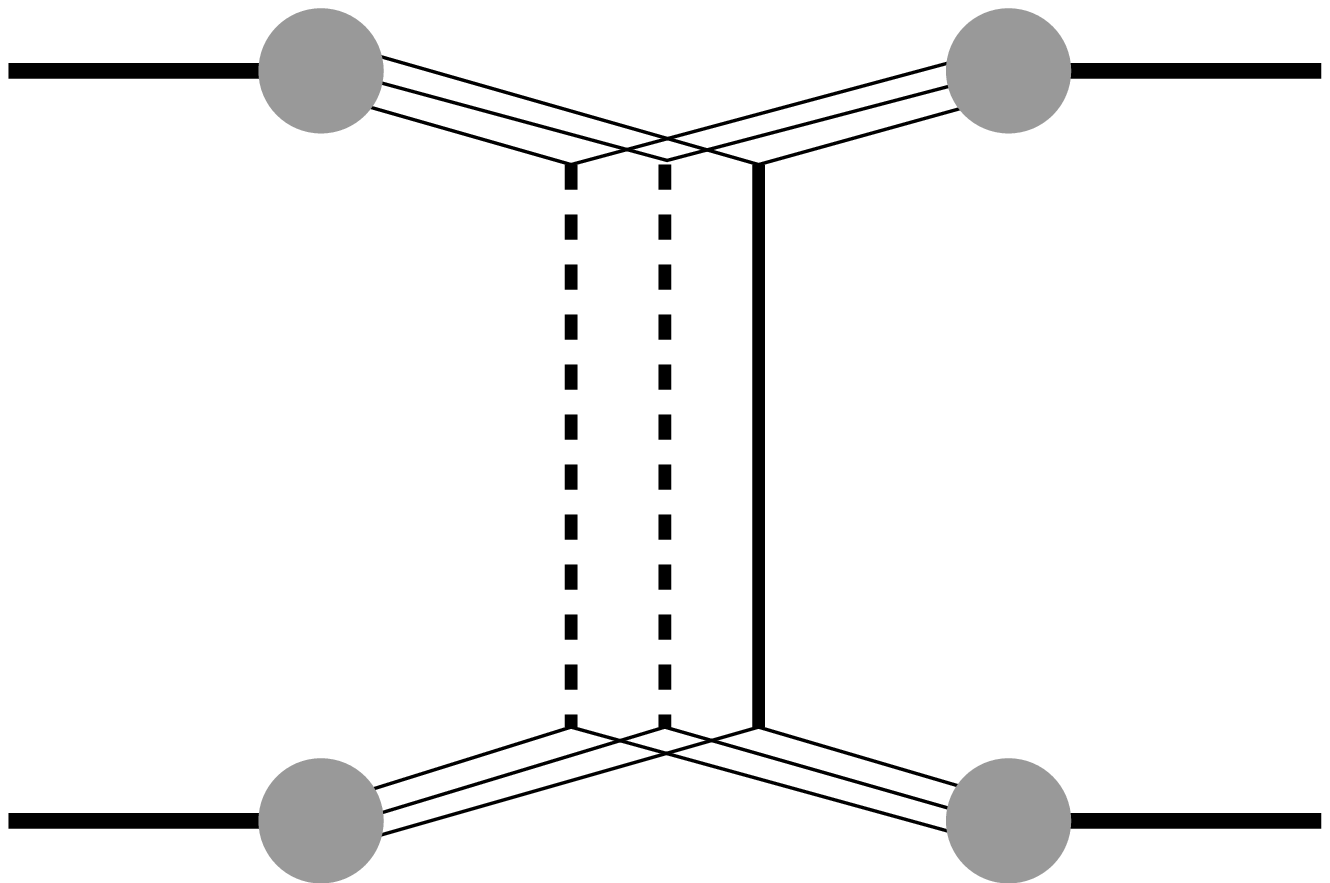}$\qquad $\includegraphics[  scale=0.35]{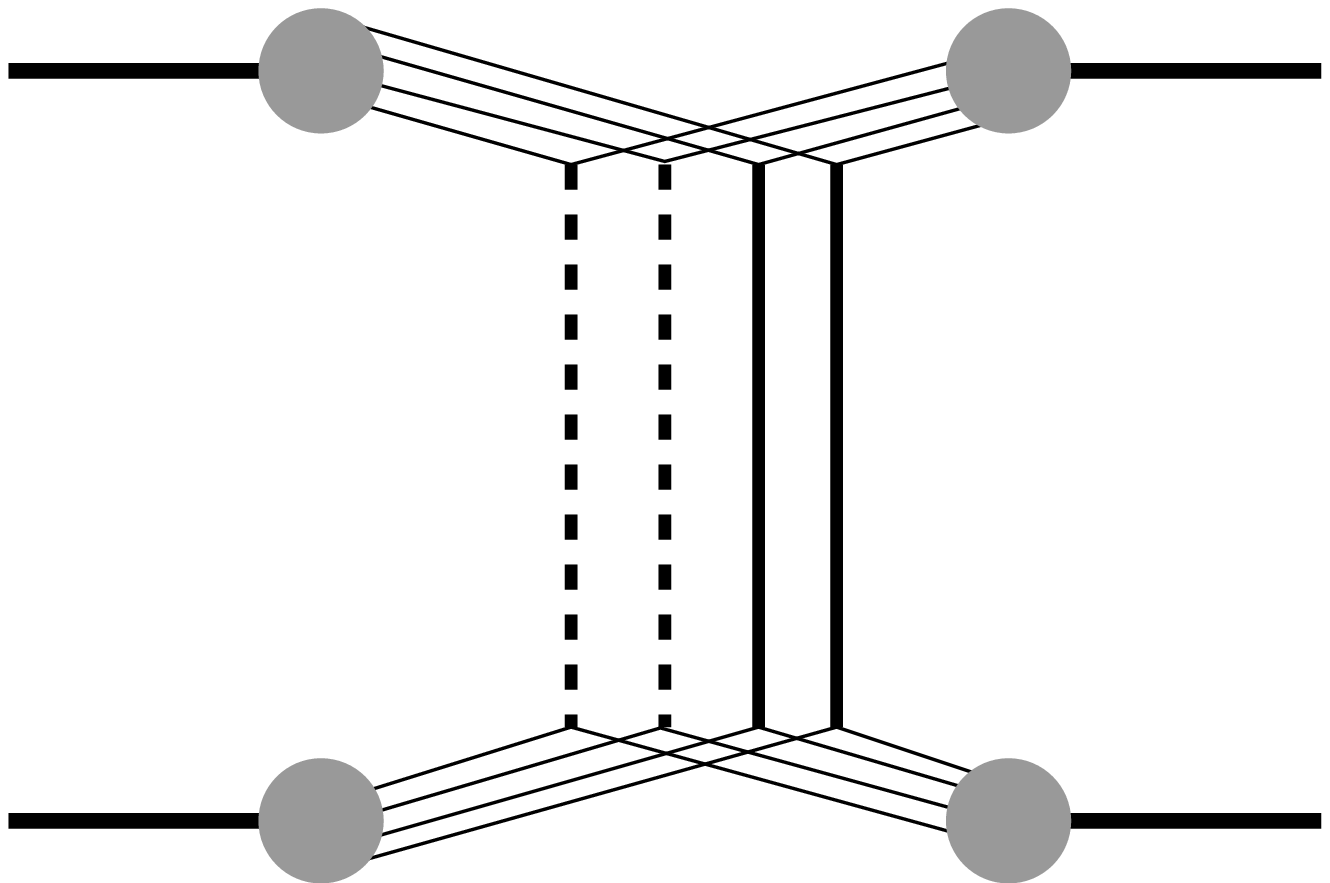}\end{center}

\caption{Class of terms corresponding to two inelastic interactions.\label{t8c}}
\end{figure}

When squaring an amplitude being a sum of many terms, not all of the
terms interfere -- only those which correspond to the same final state.
For example, a single inelastic interaction does not interfere with
a double inelastic interaction, whereas all the contributions with
exactly on inelastic interaction interfere. So considering a squared
amplitude, one may group terms together representing the same final
state. In our pictorial language, this means that all diagrams with
one dashed line, representing the same final state, may be considered
to form a class, characterized by $m=1$ -- one dashed line ( one
elementary cut) -- and the light cone momenta $x^{+}$ and $x^{-}$
attached to the dashed line (defining energy and momentum of the Pomeron).
In fig. \ref{t7c}, we show several diagrams belonging to this class,
in fig. \ref{t8c}, we show the diagrams belonging to the class of
two inelastic interactions, characterized by $m=2$ and four light-cone
momenta $x_{1}^{+}$, $x_{1}^{-}$, $x_{2}^{+}$, $x_{2}^{-}$.%
Generalizing these considerations, we may group all contributions
with $m$ inelastic interactions ($m$ dashed lines = $m$ cut Pomerons)
into a class characterized by the variable\[
K=\{m,x_{1}^{+},x_{1}^{-},\cdots ,x_{m}^{+},x_{m}^{-}\}.\]
 We then sum all the terms in a class $K$,\[
\Omega (K)=\sum \{\mathrm{all}\, \mathrm{terms}\, \mathrm{in}\, \mathrm{class}\, K\}.\]
The cross section is then simply a sum over classes,\[
\sigma _{\mathrm{inel}}(s)=\sum _{K\neq 0}\int d^{2}b\, \Omega (K).\]
$\Omega $ depends implicitly on the energy squared $s$ and the impact
parameter $b$: $\Omega =\Omega ^{(s,b)}$. The individual terms $\int d^{2}b\, \Omega (K)$,
represent partial cross sections, since they represent distinct final
states. They are referred to as topological cross sections. 

\begin{figure}
\begin{center}\includegraphics[  scale=0.4]{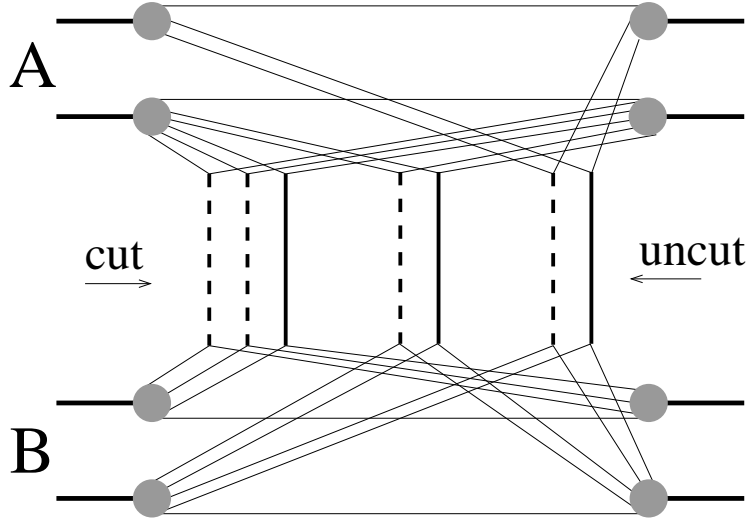}\end{center}

\caption{Nucleus-nucleus collisions: A contribution to the squared amplitude.
\label{grtpabaa2}}
\end{figure}

The above concepts are easily generalized to nucleus-nucleus scattering,
an example for a diagram representing a contribution to the squared
amplitude is shown in fig. \ref{grtpabaa2}.%
We may also define classes, which correspond to well defined final
states, in our notation a given number of dashed lines between nucleon
pairs. We may number the pairs as 1, 2, 3, ... $k$ ... , $AB$. We
define $m_{k}$ to be the number of inelastic interactions (cut Pomerons)
of the pair number $k$. The $\mu ^{\mathrm{th}}$ of these $m_{k}$
cut Pomerons is characterized by light cone momenta $x_{k\mu }^{+}$,
$x_{k\mu }^{-}$. So a class may be characterized by\[
K=\{m_{k},x_{k\mu }^{+},x_{k\mu }^{-}\}.\]
 We sum all terms in a class to obtain again a quantity called $\Omega ^{(s,b)}(K),$
such that the cross section can be written as a sum over classes\[
\sigma _{\mathrm{inel}}(s)=\sum _{K\neq 0}\int d^{2}b\, \Omega ^{(s,b)}(K),\]
as in the case of proton-proton scattering. Here, however, $b$ is
a multidimensional variable representing the impact parameter $b_{0}$
and the transverse distances $b_{k}$ of all the nucleon-nucleon pairs,\begin{equation}
b=\{b_{0},b_{1},...,b_{AB}\},\end{equation}
and $\int d^{2}b$ is a symbolic notation, meaning \begin{equation}
\int d^{2}b=\int d^{2}b_{0}\int \prod _{i=1}^{A}d^{2}b_{i}^{A}\, T_{A}(b_{i}^{A})\prod _{j=1}^{B}d^{2}b_{j}^{B}\, T_{B}(b_{j}^{B}),\end{equation}
with $A,B$ being the nuclear mass numbers and with the so-called
nuclear thickness function $T_{A}(b)$ being defined as the integral
over the nuclear density $\rho _{A(B)}$, \begin{equation}
T_{A}(b):=\int dz\, \rho _{A}(\sqrt{b^{2}+z^{2}}).\label{a.11}\end{equation}
 One can prove\[
\sum _{K}\Omega (K)=1,\]
which is a very important result justifying our interpretation of
$\Omega (K)$ to be a probability distribution for the configurations
$K$. This provides also the basis for applying Monte Carlo techniques.

The function $\Omega $ is the basis of all applications of this formalism.
It provides the basis for calculating (topological) cross sections,
but also for particle production, thus providing a consistent formalism
for all aspects of a nuclear collision.

For the sake of completeness, we provide the formula for $\Omega $,
expressed in terms of the elementary interactions $G_{k}$ (indices
$k$ express that $G$ depends on the considered pair because of ladder
splitting correction) and some vertex function $F$ \cite{nexus}:\begin{equation}
\Omega _{AB}^{(s,b)}(m,X^{+},X^{-})=\prod _{k=1}^{AB}\left\{ \frac{1}{m_{k}!}\, \prod _{\mu =1}^{m_{k}}G_{k}(x_{k,\mu }^{+},x_{k,\mu }^{-},s,b_{k})\right\} \; \Phi _{AB}\left(x^{\mathrm{proj}},x^{\mathrm{targ}},s,b\right),\label{omega-ab-bas}\end{equation}
with a function $\Phi $ representing summations over all uncut contributions
corresponding to a given cut configuration, given as\begin{eqnarray}
\Phi _{AB}\left(x^{\mathrm{proj}},x^{\mathrm{targ}},s,b\right) & = & \sum _{l_{1}}\ldots \sum _{l_{AB}}\, \label{rremnant}\\
 & \times  & \int \, \prod _{k=1}^{AB}\left\{ \prod _{\lambda =1}^{l_{k}}d\tilde{x}_{k,\lambda }^{+}d\tilde{x}_{k,\lambda }^{-}\right\} \; \prod _{k=1}^{AB}\left\{ \frac{1}{l_{k}!}\, \prod _{\lambda =1}^{l_{k}}-G_{k}(\tilde{x}_{k,\lambda }^{+},\tilde{x}_{k,\lambda }^{-},s,b_{k})\right\} \nonumber \\
 & \times  & \prod _{i=1}^{A}F_{\mathrm{remn}}\left(x_{i}^{\mathrm{proj}}-\sum _{\pi (k)=i}\tilde{x}_{k,\lambda }^{+}\right)\prod _{j=1}^{B}F_{\mathrm{remn}}\left(x_{j}^{\mathrm{targ}}-\sum _{\tau (k)=j}\tilde{x}_{k,\lambda }^{-}\right).\nonumber 
\end{eqnarray}
The different arguments are defined as\begin{eqnarray}
X^{+} & = & \left\{ x_{k,\mu }^{+}\right\} ,\\
X^{-} & = & \left\{ x_{k,\mu }^{-}\right\} ,\\
x^{\mathrm{proj}} & = & \left\{ x_{i}^{\mathrm{proj}}\right\} ,\\
x^{\mathrm{targ}} & = & \left\{ x_{j}^{\mathrm{targ}}\right\} ,\\
m & = & \{m_{k}\},\label{x}
\end{eqnarray}
and\begin{eqnarray}
x_{i}^{\mathrm{proj}} & = & 1-\sum _{\pi (k)=i}x_{k,\mu \, ,}^{+}\\
x_{j}^{\mathrm{targ}} & = & 1-\sum _{\tau (k)=j}x_{k,\mu }^{-}\, .\label{x}
\end{eqnarray}
The functions $\pi (k)$ and $\tau (k)$ refer to the projectile and
the target nucleons participating in the $k^{\mathrm{th}}$ interaction
(pair $k$) and $F_{\mathrm{remn}}$ is the vertex function to the
projectile or target remnant.

\section{Appendix on Soft and Hard Elementary Interactions\label{Appendix-soft-hard}}

What are actually these $G$ functions, which represent elementary
interactions? To explain this we have to discuss some basic facts
about scattering. Let $T$ be the elastic scattering amplitude $T$
for a $2\to 2$ scattering. The 4-momenta $p$ and $p'$ are the ones
for the incoming particles , $\tilde{p}=p+q$ and $\tilde{p}'=p'-q$
the ones for the outgoing particles, and $q$ the 4-momentum transfer
in the process.  We define as usual the Mandelstam variables $s$
and $t$. Using the optical theorem, we may write the total cross
section as \begin{equation}
\sigma _{\mathrm{tot}}(s)=\frac{1}{2s}2\mathrm{Im}\, T(s,t=0).\label{eq-2-1-1}\end{equation}
 We define the Fourier transform $\tilde{T}$ of $T$ as \begin{equation}
\tilde{T}(s,b)=\frac{1}{4\pi ^{2}}\int d^{2}q_{\bot }\, e^{-i\vec{q}_{\bot }\vec{b}}\, T(s,t),\label{eq-2-1-2}\end{equation}
using $t=-q_{\bot }^{2}$, and we define $G$ as\begin{equation}
G(s,b)=\frac{1}{2s}2\mathrm{Im}\, \tilde{T}(s,b).\label{eq-2-1-3}\end{equation}
 One can easily verify that\begin{equation}
\sigma _{\mathrm{tot}}(s)=\int \! d^{2}b\; G(s,b),\label{x}\end{equation}
which allows an interpretation of $G(s,b)$ to be the probability
of an interaction at impact parameter $b$. \\

This are actually such {}``$G$ functions'' which appear in the
multiple scattering formulas. The fact that an elementary scattering
of nucleon constituents (quarks) is composed of soft and hard components
(and a semihard one, as discussed later) can be expressed as\begin{equation}
G=G_{\mathrm{soft}}+G_{\mathrm{hard}}.\end{equation}

\subsection*{Soft Interactions}

Let us first consider a soft interaction (with only small virtualities
involved). We assume that nucleon constituents (quarks, diquarks,...)
from projectile and target may interact with the result of producing
many (low $p_{t}$) hadrons. We refer to this as an elementary soft
interaction of nucleon constituents. The corresponding amplitude ($T$-matrix
element) is assumed to be of the form \begin{equation}
T_{\mathrm{soft}}\! \! \left(s,t\right)=i\, 8\pi \, s_{0}\, \gamma _{\mathrm{part}}^{2}\, \left(\frac{s}{s_{0}}\right)^{\alpha _{\mathrm{soft}}}\exp \! \left(\lambda _{\mathrm{soft}}\! \left(s/s_{0}\right)t\right),\label{tsoft}\end{equation}
\begin{equation}
\lambda _{\mathrm{soft}}\! (z)=2R_{\mathrm{part}}^{2}+\alpha '\! _{\mathrm{soft}}\ln \! z,\label{x}\end{equation}
with parameters $\alpha _{\mathrm{soft}}$, $\alpha '\! _{\mathrm{soft}}$
, $\gamma _{\mathrm{part}}$, $R_{\mathrm{part}}^{2}$, and a scale
$s_{0}=1$GeV$^{2}$. We define $G_{\mathrm{soft}}(s,b)$ to be twice
the imaginary part of the Fourier transform (with respect to $k_{\bot }=-t$)
of $T_{\mathrm{soft}}$, as discussed above. We find\begin{equation}
G_{\mathrm{soft}}(s,b)=\frac{2\gamma _{\mathrm{part}}^{2}}{\lambda _{\mathrm{soft}}\! (s/s_{0})}\left(\frac{s}{s_{0}}\right)^{\alpha _{\mathrm{soft}}-1}\exp \! \left(-\frac{b^{2}}{4\lambda _{\mathrm{soft}}\! (s/s_{0})}\right).\end{equation}
\\

\subsection*{Hard interactions}

We now proceed to the case of hard scattering , being the other extreme,
when all internal intermediate partons are characterized by large
virtualities $Q^{2}>Q_{0}^{2}$.  Here, the corresponding amplitude
$T_{\mathrm{hard}}^{jk}\! \! \left(s,t\right)$ of the scattering
of two partons with flavors $j$ and $k$ can be calculated using
the perturbative QCD techniques \cite{alt82,rey81}. In the leading
logarithmic approximation of QCD, summing up terms where each (small)
running QCD coupling constant $\alpha _{s}(Q^{2})$ appears together
with a large logarithm $\ln (Q^{2}/\lambda _{\mathrm{QCD}}^{2})$
(with $\lambda _{QCD}$ being the infrared QCD scale), and making
use of the factorization hypothesis, one obtains the contribution
of the corresponding cut diagram for $t=q^{2}=0$ as the cut parton
ladder cross section $\sigma _{\mathrm{hard}}^{jk}(s,Q_{0}^{2})$
(strictly speaking, one obtains the ladder representation for the
process only using axial gauge), where all horizontal rungs of the
ladder are the final (on-shell) partons and the virtualities of the
virtual $t$-channel partons increase from the ends of the ladder
towards the largest momentum transfer parton-parton process. We have\begin{eqnarray}
\sigma _{\mathrm{hard}}^{jk}(s,Q_{0}^{2}) & = & \frac{1}{2s}2\mathrm{Im}\, T_{\mathrm{hard}}^{jk}\! \! \left(s,t=0\right)\nonumber \\
 & = & K\, \sum _{ml}\int dx_{B}^{+}dx_{B}^{-}dp_{\bot }^{2}{\frac{d\sigma _{\mathrm{Born}}^{ml}}{dp_{\bot }^{2}}}(x_{B}^{+}x_{B}^{-}s,p_{\bot }^{2})\label{sig-jk-hard}\\
 & \times  & E_{\mathrm{QCD}}^{jm}(x_{B}^{+},Q_{0}^{2},M_{F}^{2})\, E_{\mathrm{QCD}}^{kl}(x_{B}^{-},Q_{0}^{2},M_{F}^{2})\theta \! \left(M_{F}^{2}-Q_{0}^{2}\right),\nonumber 
\end{eqnarray}
 Here $d\sigma _{\mathrm{Born}}^{ml}/dp_{\bot }^{2}$ is the differential
$2\rightarrow 2$ parton scattering cross section, $p_{\bot }^{2}$
is the parton transverse momentum in the hard process, $m,l$ and
$x_{B}^{\pm }$ are correspondingly the types and the shares of the
light cone momenta of the partons participating in the hard process,
and $M_{F}^{2}$ is the factorization scale for the process (we use
$M_{F}^{2}=p_{\perp }^{2}/4$). The `evolution function' $E_{\mathrm{QCD}}^{jm}(Q_{0}^{2},M_{F}^{2},z)$
represents the evolution of a parton cascade from the scale $Q_{0}^{2}$
to $M_{F}^{2}$, i.e.\ it gives the number density of partons of
type $m$ with the momentum share $z$ at the virtuality scale $M_{F}^{2}$,
resulted from the evolution of the initial parton $j$, taken at the
virtuality scale $Q_{0}^{2}$. The evolution function satisfies the
usual DGLAP equation \cite{lip75,gri72,alt77,dok77} with the initial
condition $E_{\mathrm{QCD}}^{jm}(Q_{0}^{2},Q_{0}^{2},z)=\delta _{m}^{j}\; \delta (1-z)$.
The factor $K\simeq 1.5$ takes effectively into account higher order
QCD corrections.

In the following we shall need to know the contribution of the uncut
parton ladder $T_{\mathrm{hard}}^{jk}(s,t)$ with some momentum transfer
$q$ along the ladder (with $t=q^{2}$). The behavior of the corresponding
amplitudes was studied in \cite{lip86} in the leading logarithmic($1/x$
) approximation of QCD. The precise form of the corresponding amplitude
is not important for our application; we just use some of the results
of \cite{lip86}, namely that one can neglect the real part of this
amplitude and that it is nearly independent on $t$, i.e. that the
slope of the hard interaction $R_{\mathrm{hard}}^{2}$ is negligible
small, i.e. compared to the soft Pomeron slope one has $R_{\mathrm{hard}}^{2}\simeq 0$.
So we parameterize $T_{\mathrm{hard}}^{jk}(s,t)$ in the region of
small $t$ as \cite{rys92}\begin{equation}
T_{\mathrm{hard}}^{jk}(s,t)=is\, \sigma _{\mathrm{hard}}^{jk}(s,Q_{0}^{2})\: \exp \left(R_{\mathrm{hard}}^{2}\, t\right),\label{t-ladder}\end{equation}
 The corresponding {}``$G$ function'' is obtained by calculating
the Fourier transform $\tilde{T}$ of $T$ and dividing by the initial
parton flux $2\hat{s}$, \begin{equation}
G_{\mathrm{hard}}^{jk}\! \left(s,b\right)=\frac{1}{2s}2\mathrm{Im}\tilde{T}_{\mathrm{hard}}^{jk}(s,b),\label{x}\end{equation}
 which gives\[
G_{\mathrm{hard}}^{jk}\left(s,b\right)=\frac{1}{8\pi ^{2}s}\int d^{2}q_{\perp }\, \exp \! \left(-i\vec{q}_{\perp }\vec{b}\right)\, 2\mathrm{Im}\, T_{\mathrm{hard}}^{jk}(s,-q_{\perp }^{2}),\]
or explicitly\begin{equation}
G_{\mathrm{hard}}^{jk}\left(s,b\right)==\sigma _{\mathrm{hard}}^{jk}\! \left(s,Q_{0}^{2}\right)\frac{1}{4\pi R_{\mathrm{hard}}^{2}}\exp \! \left(-\frac{b^{2}}{4R_{\mathrm{hard}}^{2}}\right).\end{equation}
 This is not yet the hard $G$ function we are looking for, there
is still one element missing, namely the relation between the partons
$j$ and $k$ and the incident nucleons. There are two elements relevant:
the vertex function, and the soft pre-evolution.

\subsection*{The vertex function}

The coupling of the multiple (say $n$) elementary interactions to
the nucleons is expressed via projectile and target nucleon vertex
functions, having the form\begin{equation}
F_{\mathrm{remn}}^{N}\! \left(1-\sum _{k=1}^{n}x_{k}\right)\; \exp \! \left(-R_{N}^{2}\sum _{k=1}^{n}\! q_{k_{\perp }}^{2}\right)\; \prod _{k=1}^{n}\! F_{\mathrm{part}}^{N}(x_{k})\, ,\end{equation}
with\begin{equation}
F_{\mathrm{part}}^{N}(x)=\gamma _{N}\, x^{-\alpha _{\mathrm{part}}},\label{f-part}\end{equation}
\begin{equation}
F_{\mathrm{remn}}^{N}\! (x)=x^{\alpha _{\mathrm{remn}}^{N}}\, \Theta \left(x\right)\, \Theta \left(1-x\right),\label{f-remn}\end{equation}
with parameters $R_{N}^{2}$, $\gamma _{N}$, $\alpha _{\mathrm{part}}$,
$\alpha _{\mathrm{remn}}^{N}$. All parameters with an {}``$N$''
sub(super)script are specific for projectile/target nucleons and may
be different for other kinds of hadrons, like pions, kaons. The arguments
$x_{k}$ and $q_{k_{\perp }}^{2}$ are respectively the light cone
momentum fractions and squared transverse momenta of the nucleon constituents
participating in an interaction. All factors apart of $F_{\mathrm{remn}}$
can be absorbed into the elementary $G$ functions.

\subsection*{Soft pre-evolution}

In case of sea quarks and gluons being at the end of a parton ladder,
the momentum share $x_{1}$ of the {}``first'' parton is typically
very small, leading to an object with a large mass of the order $Q_{0}^{2}/x_{1}$
between the parton and the proton \cite{don94}. Microscopically,
such 'slow' partons with $x_{1}\ll 1$ appear as a result of a long
non-perturbative parton cascade, where each individual parton branching
is characterized by a small momentum transfer squared $Q^{2}<Q_{0}^{2}$
and nearly equal partition of the parent parton light cone momentum
\cite{afs62,bak76}. When calculating proton structure functions or
high-$p_{t}$ jet production cross sections, this non-perturbative
contribution is usually included into parameterized initial parton
momentum distributions at $Q^{2}=Q_{0}^{2}$. However, the description
of inelastic hadronic interactions requires to treat it explicitly
in order to account for secondary particles produced during such non-perturbative
parton pre-evolution, and to describe correctly energy-momentum sharing
between multiple elementary scatterings. As the underlying dynamics
appears to be identical to the one of soft parton-parton scattering
considered above, we treat this soft pre-evolution as the usual soft
emission. We account for this by introducing {}``soft pre-evolutions
functions'' $E_{\mathrm{soft}}$, as discussed in more detail later.

\subsection*{The complete hard contribution}

The complete $G$ function, including soft pre-evolution and vertex
contributions, is given as\begin{eqnarray}
G_{\mathrm{hard}}\left(x^{+},x^{-},s,b\right) & = & F_{\mathrm{part}}^{N}(x^{+})\, F_{\mathrm{part}}^{N}(x^{-})\, \\
 &  & \nonumber \\
 &  & \qquad \sum _{jk}\int _{0}^{1}dz^{+}dz^{-}E_{\mathrm{soft}}^{j}\left(z^{+}\right)\, E_{\mathrm{soft}}^{k}\left(z^{-}\right)\, \sigma _{\mathrm{hard}}^{jk}(z^{+}z^{-}x^{+}x^{-}s,Q_{0}^{2})\nonumber \\
 &  & \qquad \frac{1}{4\pi \, \lambda _{NN}(1/(z^{+}z^{-}))}\exp \! \left(-\frac{b^{2}}{4\lambda _{NN}\! \left(1/(z^{+}z^{-})\right)}\right)\nonumber 
\end{eqnarray}
with $\lambda _{NN}(\xi )=2R_{N}^{2}+\alpha '_{\mathrm{soft}}\ln \xi $.
The variables $x^{+}$, $x^{-}$ refer to the soft nucleon constituents
participating in an elementary interaction. The variables $z^{+}$,
$z^{-}$ refer to the first hard partons, the ladder ends. The function
$E_{\mathrm{soft}}^{k}$ has two contributions, \begin{equation}
E_{\mathrm{soft}}^{k}(z)=E_{\mathrm{soft}(0)}^{k}(z)+E_{\mathrm{soft}(1)}^{k}(z).\end{equation}
The first term, representing a soft {}``pre-evolution'', is given
as ~Im$T_{\mathrm{soft}}(s_{0}/z,t=0)$, up to some coupling modifications,
explicitely\begin{equation}
E_{\mathrm{soft}(0)}^{q}(z)=z^{-\alpha _{\mathrm{soft}}}\, \tilde{E}_{\mathrm{soft}(0)}^{q}(z),\quad E_{\mathrm{soft}(0)}^{g}(z)=z^{-\alpha _{\mathrm{soft}}}\, \tilde{E}_{\mathrm{soft}(0)}^{g}(z),\end{equation}
with\begin{equation}
\tilde{E}_{\mathrm{soft}(0)}^{q}(z)=\gamma _{\mathrm{soft}}\, w\, \hat{E}_{\mathrm{soft}(0)}^{q}(z),\quad \tilde{E}_{\mathrm{soft}(0)}^{g}(z)=\gamma _{\mathrm{soft}}\, (1-w)\, (1-z)^{\beta },\end{equation}
with $\gamma _{\mathrm{soft}}=8\pi s_{0}\gamma _{\mathrm{part}}\tilde{\gamma }$,
and\begin{equation}
\hat{E}_{\mathrm{soft}(0)}^{q}(z)=\int _{x}^{1}d\xi \, \xi ^{\delta _{s}}\, P(\xi )\, (1-\frac{z}{\xi })^{\beta },\end{equation}
with some parameters $\tilde{\gamma }$ and $\beta $.

The second term, representing a direct coupling to a valence quark,
and which has therefore only a quark contribution, can be written
as \begin{equation}
E_{\mathrm{soft}(1)}^{q}(x,z)=z^{-\alpha _{\mathrm{soft}}}\, \tilde{E}_{\mathrm{soft}(1)}^{q}(z\, x,x),\end{equation}
with\begin{eqnarray}
\tilde{E}_{\mathrm{soft}(1)}^{q}(x_{q},x) & = & x\, \frac{1}{\gamma _{h}}x^{\alpha _{\mathrm{part}}}\frac{\Gamma (2+\alpha _{\mathrm{remn}}-\alpha _{\mathrm{I}\! \mathrm{R}})}{\Gamma (1+\alpha _{\mathrm{remn}})\Gamma (1-\alpha _{\mathrm{I}\! \mathrm{R}})}\, \\
 &  & \nonumber \\
 &  & \qquad \qquad q_{\mathrm{val}}(x_{q})(1-x_{q})^{-1-\alpha _{\mathrm{remn}}+\alpha _{\mathrm{I}\! \mathrm{R}}}(x-x_{q})^{-\alpha _{\mathrm{I}\! \mathrm{R}}}.\nonumber 
\end{eqnarray}
So this function depends also on the momentum fraction $x$. We neglected
the small hard scattering slope $R_{\mathrm{hard}}^{2}$ compared
to the Pomeron slope $\lambda _{\mathrm{soft}}$. 

We call $E_{\mathrm{soft}}$ also the {}`` soft evolution'', to
indicate that we consider this as simply  a continuation of the QCD
evolution, however, in a region where perturbative  techniques do
not apply any more. $E_{\mathrm{soft}}^{j}\left(z\right)$ has the
meaning of the momentum distribution of parton $j$ in the soft piece.
More details can be found in \cite{nexus}.

\section{Appendix on Parameterizations of G Functions\label{Appendix-Parameterizations}}

The calculations of partial cross sections in our multiple scattering
theory involve high dimensional integrations, excluding numerical
techniques. Fortunately, the numerically determined $G$ function
(the sum of soft and hard) can to a high precision be expressed as 

\begin{equation}
G(x^{+},x^{-},s,b)=(x^{+}x^{-})^{-\alpha _{\mathrm{part}}}\sum _{i=1}^{2}\alpha _{G_{i}}\, s^{\beta _{G_{i}}}\, (x^{+}x^{-})^{\beta '_{G_{i}}}\, (sx^{+}x^{-})^{\gamma _{G_{i}}b^{2}}\exp (-\frac{b^{2}}{\delta _{G_{i}}}).\end{equation}
This form is inspired by the fact that $G$ is a sum of two contributions,
soft ($i=1)$ and hard ($i=2$), with quite different exponents $\beta '_{G_{i}}$.
The fit parameters are, however, chosen such that the sum of both,
soft and hard, is best reproduced. The above form provides in fact
excellent fits of the $x^{+}$, $x^{-}$, and also the $b$ dependence
of $G$. Concerning this $b$ dependence, one has\begin{equation}
(sx^{+}x^{-})^{\gamma _{G_{i}}b^{2}}\exp (-\frac{b^{2}}{\delta _{G_{i}}})\approx \exp (-\frac{b^{2}}{\lambda _{G_{i}}}),\end{equation}
with\begin{equation}
\lambda _{G_{i}}=\delta _{G_{i}}+\delta _{G_{i}}^{2}\gamma _{G_{i}}\ln (sx^{+}x^{-}),\end{equation}
indicating that the width of the $b$ dependence increases logarithmically
with energy. 

Using the analytical form of $G$, many of the multidimensional integrals
can be done analytically. For a given $s$ and $b$, one may write\begin{equation}
G(x^{+},x^{-},s,b)=\sum _{i=1}^{2}\alpha _{i}\, (x^{+}x^{-})^{\beta _{i}},\end{equation}
with $s$ and $b$ dependent coefficients $\alpha _{i},\: \beta _{i}$.
We also use the notation $\beta _{S}\equiv \beta _{1}$ and $\beta _{H}\equiv \beta _{2}.$
Taking into account screening via the exponents $\varepsilon _{S}\equiv \varepsilon _{1}$
and $\varepsilon _{H}\equiv \varepsilon _{2}$ (see section \ref{sec:Realization-of-Ladder}),
we have\begin{equation}
G(x^{+},x^{-},s,b)=\sum _{i=1}^{2}\alpha _{i}\, (x^{+})^{\beta _{i}+\varepsilon _{i}(Z_{P})}\, (x^{-})^{\beta _{i}+\varepsilon _{i}(Z_{T})}.\end{equation}

The first task is the calculation of the function $\Phi $, representing
summations over all uncut contributions corresponding to a given cut
configuration, which we need to know to be able to calculate partial
cross section. For nucleon-nucleon, one obtains\begin{equation}
\Phi _{NN}(x^{+},x^{-},s,b)=(x^{+}x^{-})^{\alpha _{\mathrm{remn}}}\sum _{i=0}^{\infty }\sum _{j=0}^{\infty }\frac{(a_{1})^{i}}{i!}\, \frac{(a_{2})^{j}}{j!}\, g(i\tilde{\beta }_{1}+j\tilde{\beta }_{2})\, g(i\tilde{\beta }'_{1}+j\tilde{\beta }'_{2}),\end{equation}
with $a_{i}=-\alpha _{i}\Gamma (\tilde{\beta }_{i})\Gamma (\tilde{\beta '}_{i})(x^{+})^{\tilde{\beta }_{i}}(x^{-})^{\tilde{\beta '}_{i}}$,
and $\tilde{\beta }_{i}=\beta _{i}+\varepsilon _{i}(Z_{P})+1$, $\tilde{\beta }'_{i}=\beta _{i}+\varepsilon _{i}(Z_{T})+1$.
The function $g$ is given as\[
g(x)=\frac{\Gamma (1+\alpha _{\mathrm{remn}})}{\Gamma (1+\alpha _{\mathrm{remn}}+x)}\, .\]
$\Phi _{NN}$ is almost an exponential function. So to regularize
the theory, we replace $g(i\tilde{\beta }_{1}+j\tilde{\beta }_{2})$
by $g(\tilde{\beta }_{1})^{i}g(\tilde{\beta }_{2})^{j}$, to get an
exponential-type function. It is not precisely the same, but we take
this freedom, because even small deviations from an exponential form
of $\Phi $ may lead to drastic consequences (big oscillations), not
at all being physical. All these considerations can be generalized
to nucleus-nucleus $(AB)$ collisions, where we have to evaluate $\Phi _{AB}$.
One gets\begin{eqnarray}
\Phi _{AB}(x^{\mathrm{proj}},x^{\mathrm{targ}},s,b) & = & \prod _{k=1}^{AB}\exp \left(-\tilde{G}_{k}(x_{\pi (k)}^{+},x_{\tau (k)}^{-},s,b_{k})\right)\\
 & \times  & \prod _{i=1}^{A}(x_{i}^{+})^{\alpha _{\mathrm{remn}}}\, \Theta (x_{i}^{+})\, \Theta (1-x_{i}^{+})\prod _{j=1}^{B}(x_{j}^{-})^{\alpha _{\mathrm{remn}}}\, \Theta (x_{j}^{-})\, \Theta (1-x_{j}^{-}),\nonumber 
\end{eqnarray}
with

\begin{equation}
\tilde{G}_{k}(x^{+},x^{-},s,b)=\sum _{i=1}^{2}\tilde{\alpha }_{i}^{k}\, (x^{+})^{\tilde{\beta }_{i}^{k}}\, (x^{-})^{\tilde{\beta '}_{i}^{k}},\label{G-tilde}\end{equation}
 with\begin{eqnarray}
\tilde{\alpha }_{i}^{k} & = & \alpha _{i}^{k}\, \frac{\Gamma (\tilde{\beta }_{i}^{k})\Gamma (1+\alpha _{\mathrm{remn}})}{\Gamma (1+\alpha _{\mathrm{remn}}+\tilde{\beta }_{i}^{k})}\, \frac{\Gamma (\tilde{\beta '}_{i}^{k})\Gamma (1+\alpha _{\mathrm{remn}})}{\Gamma (1+\alpha _{\mathrm{remn}}+\tilde{\beta '}_{i}^{k})},\\
\tilde{\beta }_{i}^{k} & = & \beta _{i}^{k}+\varepsilon _{i}(Z_{P})+1,\\
\tilde{\beta '}_{i}^{k} & = & \beta _{i}^{k}+\varepsilon _{i}(Z_{T})+1,
\end{eqnarray}
 with\begin{eqnarray}
\alpha _{i}^{k} & = & \alpha _{G_{i}}\exp (-\frac{b_{k}^{2}}{\delta _{G_{i}}})\, s^{\beta _{G_{i}}+\gamma _{G_{i}}b_{k}^{2}},\\
\beta _{i}^{k} & = & \beta '_{G_{i}}+\gamma _{G_{i}}b_{k}^{2}-\alpha _{\mathrm{part}}.\\
 &  & \nonumber 
\end{eqnarray}

More details can be found in \cite{nexus}.

\section{Appendix on Markov Chain Techniques\label{Appendix-Markov}}

As discussed earlier, the function $\Omega $ (the integrand of partial
cross sections) can be interpreted as probability distribution, which
allows to apply the Monte Carlo technique. So we want to generate
configurations $\{m,X^{+},X^{-}\}$ according to\begin{equation}
\Omega _{AB}^{(s,b)}(m,X^{+},X^{-})=\prod _{k=1}^{AB}\left\{ \frac{1}{m_{k}!}\, \prod _{\mu =1}^{m_{k}}G_{k}(x_{k,\mu }^{+},x_{k,\mu }^{-},s,b_{k})\right\} \; \Phi _{AB}\left(X^{\mathrm{proj}},X^{\mathrm{targ}},s,b\right),\label{x}\end{equation}
with\begin{equation}
b=\{b_{k}\},\; m=\{m_{k}\},\; X^{+}=\left\{ x_{k,\mu }^{+}\right\} ,\; X^{-}=\left\{ x_{k,\mu }^{-}\right\} ,\; X^{\mathrm{proj}}=\left\{ x_{i}^{\mathrm{proj}}\right\} ,\; X^{\mathrm{targ}}=\left\{ x_{j}^{\mathrm{targ}}\right\} ,\label{x}\end{equation}
Here, $b_{0}$ is the impact parameter between the two nuclei, and
$b_{k}$ is the transverse distance between the nucleons of $k^{th}$
pair, $m_{k}$ is the number of elementary inelastic interaction for
the nucleon-nucleon pair $k$, $x_{k,\mu }^{+}$ and $x_{k,\mu }^{-}$
are the light cone momenta of the $\mu ^{\mathrm{th}}$ interaction
of the pair $k$. The arguments of $\Phi $ are the number of inelastic
interactions $m$ and the momentum fractions of projectile and target
remnants,

\begin{equation}
x_{i}^{\mathrm{proj}}=1-\sum _{\pi (k)=i}x_{k,\mu \, ,}^{+}\quad x_{j}^{\mathrm{targ}}=1-\sum _{\tau (k)=j}x_{k,\mu }^{-}\, ,\label{x}\end{equation}
where $\pi (k)$ and $\tau (k)$ point to the remnants linked to the
$k^{\mathrm{th}}$ interaction. In the following, we perform the analysis
for given $s$ and $b=(b_{0},b_{1},...,b_{AB})$, so we do not write
these variables explicitly. Furthermore, we suppress the index $AB$.
For any given configuration the function $\Omega $ can be easily
calculated, using the techniques developed earlier. 

\begin{figure}
\begin{center}\includegraphics[  scale=0.4]{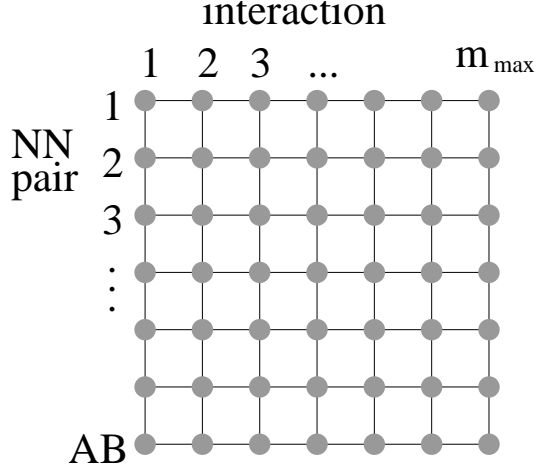}\end{center}

\caption{The interaction lattice.\label{lattice}}
\end{figure}

Since $\Omega (m,X^{+},X^{-})$ is a high-dimensional and nontrivial
probability distribution, the only way to proceed amounts to employing
dynamical Monte Carlo methods, well known in statistical and solid
state physics. We first need to choose the appropriate framework for
our analysis. So we translate our problem into the language of spin
systems \cite{hla98}: we number all nucleon pairs as 1, 2, ..., $AB$
and for each nucleon pair $k$ the possible elementary interactions
as 1,2, ..., $m_{k\cdot }$ Let $m_{\mathrm{max}}$ be the maximum
number of elementary interactions per nucleon pair one may imagine.
We now consider a two dimensional lattice with $AB$ lines and $m_{\mathrm{max}}$
columns, see fig.\ \ref{lattice}. Lattice sites are occupied $\left(=1\right)$
or empty $\left(=0\right)$, representing an elementary interaction
(1) or the case of no interaction (0), for the $k^{th}$ pair.%
In order to represent $m_{k}$ elementary interactions for the pair
$k$, we need $m_{k}$ occupied cells (1's) in the $k^{th}$ line.
A line containing only empty cells (0's) represents a pair without
interaction. Any possible interaction may be represented by this {}``interaction
matrix'' $M$ with elements\begin{equation}
m_{kn}\in \left\{ 0,1\right\} .\label{x}\end{equation}
 Such an {}``interaction configuration'' is exactly equivalent to
a spin configuration of the Ising model. Unfortunately the situation
is somewhat more complicated in case of nuclear collisions: we need
to consider the energy available for each elementary interaction,
represented via the momentum fractions $x_{kn}^{+}$ and $x_{kn}^{-}$.
So we have a {}``generalized'' matrix $K=(M,X^{+},X^{-}),$ representing
an interaction configuration. Since there are\begin{equation}
c=\prod _{k=1}^{AB}\frac{m_{\mathrm{max}}!}{m_{k}!(m_{\mathrm{max}}-m_{k})!}\label{x}\end{equation}
 configurations $(M,X^{+},X^{-})$ representing the same configuration
$(m,X^{+},X^{-})$, the weight for the former is $c^{-1}$ times the
weight for the latter, so we obtain the following probability distribution
for $K=(M,X^{+},X^{-})$: \begin{equation}
\Omega (K)=\prod _{k=1}^{AB}\left\{ \frac{(m_{\mathrm{max}}-m_{k})!}{m_{\mathrm{max}}!}\, \prod _{n=1}^{m_{\max }}\delta _{m_{kn}1}\, G_{k}(x_{kn}^{+},x_{kn}^{-},s,b)\right\} \; \Phi _{AB}\left(X^{\mathrm{proj}},X^{\mathrm{targ}},s,b\right).\label{x}\end{equation}
 In order to generate $K$ according to the given distribution $\Omega \left(K\right)$,
defined earlier, we construct a chain of configurations $K^{(t)}$
such that the final configurations $K^{\left(t_{\mathrm{max}}\right)}$
are distributed according to the probability distribution $\Omega \left(K\right)$,
if possible for a $t_{\mathrm{max}}$ not too large! Let us discuss
how to obtain a new configuration $K^{(t+1)}=L$ from a given configuration
$K^{(t)}=K$. We use Metropolis' Ansatz for the transition probability\begin{equation}
p(K,L)=prob\left(K^{\left(t+1\right)}=L\: \right|\left.K^{\left(t\right)}=K\right)\label{x}\end{equation}
 as a product of a proposition matrix $w(K,L)$ and an acceptance
matrix $u(K,L)$, where we use \begin{equation}
u(K,L)=\min \left(\frac{\Omega (L)}{\Omega (K)}\frac{w(L,K)}{w(K,L)},1\right),\label{x}\end{equation}
 in order to assure detailed balance. We use \begin{equation}
w(K,L)=\left\{ \begin{array}{ccc}
 \Omega _{0}(L) & \mathrm{if} & d(K,L)\leq 1\\
 0 & \mathrm{otherwise} & \end{array}
\right.,\label{x}\end{equation}
 where $d(K,L)$ is the number of lattice sites being different in
$L$ compared to $K,$ and where $\Omega _{0}$ is defined by the
same formulas as $\Omega $ with one exception : $\Phi $ is replaced
by $\prod _{k=1}^{AB}\prod _{n=1}^{m_{\max }}\delta _{m_{kn}1}\left(1-x_{kn}^{+}\right)\left(1-x_{kn}^{-}\right)$.
So we get\begin{eqnarray}
\Omega _{0}(L) & \sim  & \prod _{k=1}^{AB}\big \{\, (m_{\mathrm{max}}-m_{k})!\label{omega0}\\
 & \times  & \prod _{n=1}^{m_{\max }}\delta _{m_{kn}1}\, G(x_{kn}^{+},x_{kn}^{-},s,b_{k})\left(1-x_{kn}^{+}\right)\left(1-x_{kn}^{-}\right)\, \big \}.\nonumber 
\end{eqnarray}
The above definition of $w(K,L)$ may be realized by the following
algorithm:

\begin{itemize}
\item choose randomly a lattice site $(k,n)$,
\item propose a new matrix element $(m_{kn},x_{kn}^{+},x_{k,n}^{-})$ according
to the probability distribution $\rho (m_{kn},x_{kn}^{+},x_{k,n}^{-})$,
\end{itemize}
where we are going to derive the form of $\rho $ in the following.
From eq.\ (\ref{omega0}), we know that $\rho $ should be of the
form\begin{equation}
\rho (m,x^{+},x^{-})\sim m_{0}!\, \left\{ \begin{array}{ccc}
 G(x^{+},x^{-},s,b)\left(1-x^{+}\right)\left(1-x^{-}\right) & \mathrm{if} & m=1\\
 1 & \mathrm{if} & m=0\end{array}
\right.,\label{x}\end{equation}
where $m_{0}=m_{\mathrm{max}}-m$ is the number of zeros in the row
$k$. Let us define $\bar{m}_{0}$ as the number of zeros (empty cells)
in the row $k$ not counting the current site $(k,\mu )$. Then the
factor $m_{0}!$ is given as $\bar{m}_{0}!$ in case of $m\neq 0$
and as $\bar{m}_{0}!(\bar{m}_{0}+1)$ in case of $m=0$, and we obtain\begin{equation}
\rho (m,x^{+},x^{-})\sim (\bar{m}_{0}+1)\delta _{m0}+G(x^{+},x^{-},s,b)\left(1-x^{+}\right)\left(1-x^{-}\right)\delta _{m1}.\label{x}\end{equation}
Properly normalized, we obtain\begin{equation}
\rho (m,x^{+},x^{-})=p_{0}\, \delta _{m0}+(1-p_{0})\, \frac{G(x^{+},x^{-},s,b)\left(1-x^{+}\right)\left(1-x^{-}\right)}{\chi }\, \delta _{m1},\label{x}\end{equation}
where the probability $p_{0}$ of proposing no interaction is given
as\begin{equation}
p_{0}=\frac{\bar{m}_{0}+1}{\bar{m}_{0}+1+\chi (s,b)},\label{x}\end{equation}
 with $\chi $ being obtained by integrating $G(1-x^{+})(1-x^{-})$
over $x^{+}$ and $x^{-}$, \begin{equation}
\chi (s,b)=\int _{0}^{1}dx^{+}dx^{-}G(x^{+},x^{-},s,b)\left(1-x^{+}\right)\left(1-x^{-}\right).\label{x}\end{equation}
Having proposed a new configuration $L$, which amounts to generating
the values $m_{kn},\, x_{kn}^{+},\, x_{kn}^{-}$ for a randomly chosen
lattice site as described above, we accept this proposal with the
probability\begin{equation}
u(K,L)=\min \left(z_{1}z_{2},1\right),\label{x}\end{equation}
with \begin{equation}
z_{1}=\frac{\Omega (L)}{\Omega (K)},\quad z_{2}=\frac{w(L,K)}{w(K,L)}.\label{x}\end{equation}
Since $K$ and $L$ differ in at most one lattice site, say $(k,n)$,
we do not need to evaluate the full formula for the distribution $\Omega $
to calculate $z_{1}$, we rather calculate \begin{equation}
z_{1}=\frac{\Omega ^{kn}(L)}{\Omega ^{kn}(K)},\label{x}\end{equation}
with\begin{eqnarray}
\Omega ^{kn}(K) & = & \left((\bar{m}_{0}+1)\delta _{m_{kn}0}+\delta _{m_{kn}1}\, G_{k}(x_{kn}^{+},x_{kn}^{-},s,b_{k})\right)\nonumber \\
 & \times  & \exp \left(-\sum _{l\, \mathrm{linked}\, \mathrm{to}\, k}\tilde{G}_{l}(x_{\pi (l)}^{\mathrm{proj}},x_{\tau (l)}^{\mathrm{targ}},s,b_{l})\right)\\
 & \times  & (x_{\pi (k)}^{\mathrm{proj}})^{\alpha _{\mathrm{remn}}}\Theta (x_{\pi (k)}^{\mathrm{proj}})\Theta (1-x_{\pi (k)}^{\mathrm{proj}})\, (x_{\tau (k)}^{\mathrm{targ}})^{\alpha _{\mathrm{remn}}}\Theta (x_{\tau (k)}^{\mathrm{targ}})\Theta (1-x_{\tau (k)}^{\mathrm{targ}}),\nonumber 
\end{eqnarray}
which is technically quite easy. Our final task is the calculation
of the asymmetry $z_{2}$. In many applications of the Markov chain
method one uses symmetric proposal matrices, in which case this factor
is simply one. This is not the case here. We have\begin{equation}
z_{2}=\frac{\Omega _{0}(K)}{\Omega _{0}(L)}=\frac{\Omega _{0}^{kn}(K)}{\Omega _{0}^{kn}(L)},\label{x}\end{equation}
with\begin{equation}
\Omega _{0}^{kn}(K)=\rho (m_{kn},x_{kn}^{+},x_{kn}^{-}),\label{x}\end{equation}
which is also easily calculated. So we accept the proposal $L$ with
the probability $\min (z_{1}z_{2},1)$, in which case we have $K^{(t+1)}=L$,
otherwise we keep the old configuration $K$, which means $K^{(t+1)}=K$.

\newpage
~

\end{document}